\documentclass[11pt]{article}
\usepackage[margin=1in]{geometry}
\usepackage{xr}

\RequirePackage{amsmath,amsfonts,amssymb,amsthm}
\usepackage{longtable}
\usepackage{threeparttable}
\usepackage[caption=false]{subfig}
\usepackage[doublespacing]{setspace}
\usepackage{pdflscape}

\usepackage[longnamesfirst,sort]{natbib}
\bibpunct[, ]{(}{)}{;}{a}{,}{,}%

\usepackage{xspace}

\usepackage{placeins}

\usepackage[utf8]{inputenc} 
\usepackage{bm}
\usepackage[hidelinks]{hyperref}
\usepackage{graphicx}
\usepackage{mathtools}
\usepackage{makecell}
\usepackage{booktabs}
\usepackage{multirow}
\usepackage{numdef}
\usepackage{rotating}

\usepackage[dvipsnames,table]{xcolor}

\definecolor{lightgray}{gray}{0.9}

\newcommand{\rd}[1]{\textcolor{BrickRed}{#1}}

\newcommand{\bx}{\bm{x}}

\newcommand{\bxs}{\bx}
\newcommand{\bxy}{\bm{g}(\bx)}
\newcommand{\bxyt}{\bxy^\prime}
\newcommand{\bxsi}{\bxs_i}
\newcommand{\bxsit}{\bm{\tilde{x}^S}_i}
\newcommand{\bxsitp}{\bm{\tilde{x}^{S\prime}}_i}
\newcommand{\EE}{\mathbb{E}}
\newcommand{\pp}{e(\bxs)}
\newcommand{\ppi}{e(\bxsi)}
\newcommand{\ppip}[1]{e_{#1}(\bxsi)}

\newcommand{\hppi}{\hat{e}(\bxsi)}
\newcommand{\ppf}[2]{e_{#1}\left(#2\right)}
\newcommand{\yt}{Y_T}
\newcommand{\yc}{Y_C}
\newcommand{\yti}{Y_{Ti}}
\newcommand{\yci}{Y_{Ci}}
\newcommand{\sti}{S_{Ti}}
\newcommand{\sci}{S_{Ci}}
\newcommand{\st}{S_T}
\num\newcommand{\eff0}{\tau^0}
\num\newcommand{\eff1}{\tau^1}
\num\newcommand{\mut1}{\mu_T^1}
\num\newcommand{\mut0}{\mu_T^0}
\num\newcommand{\muc1}{\mu_C^1}
\num\newcommand{\muc0}{\mu_C^0}
\num\newcommand{\heff0}{\hat{\tau}^0}
\num\newcommand{\heff1}{\hat{\tau}^1}
\newcommand{\mucs}{\mu_C^s}
\newcommand{\muts}{\mu_T^s}
\newcommand{\effs}{\tau^s}
\newcommand{\ri}{R_i}
\newcommand{\blam}{\bm{\Lambda}}

\newcommand{\geepers}{\textsc{geepers}\xspace}
\newcommand{\psw}{\textsc{psw}\xspace}
\newcommand{\pmm}{\textsc{pmm}\xspace}

\newcommand\independent{\protect\mathpalette{\protect\independenT}{\perp}}
\def\independenT#1#2{\mathrel{\rlap{$#1#2$}\mkern2mu{#1#2}}}

\theoremstyle{plain}
\newtheorem{prop}{Proposition}
\newtheorem{lemma}{Lemma}

\theoremstyle{remark}
 \newtheorem{ass}{Assumption}

\newtheorem{assumptionalt}{Assumption}[ass]

\newenvironment{assp}[1]{
  \renewcommand\theassumptionalt{#1}
  \assumptionalt
}{\endassumptionalt}

\begin{document}


\title{GEEPERs: Principal Stratification using Principal Scores and Stacked Estimating Equations}


 \author{
 Adam C. Sales\thanks{CONTACT Adam C. Sales. Email: asales@wpi.edu}, Kirk P. Vanacore, and Erin R. Ottmar
}

\date{}

\maketitle

\begin{abstract}
Principal stratification is a framework for making sense of causal effects conditioned on variables that may themselves have been affected by the treatment. For instance, in an evaluation of an educational intervention, some subjects in the treatment group may not fully utilize the intervention, and researchers may be interested in how this subgroup is affected. Most principal stratification estimators rely on strong structural or modeling assumptions and often require advanced statistical training to fit and evaluate, making them inaccessible to many applied researchers. In this paper, we introduce a new principal effect estimator for one-way noncompliance based on a binary indicator. Estimates may be computed using conventional regression methods (though the standard errors require a specialized sandwich estimator) and do not rely on distributional assumptions. We present a simulation study that demonstrates the novel method's greater robustness compared to popular alternatives and illustrate the method through a real-data analysis.
\end{abstract}

\clearpage

\section{Introduction}
Participants in educational randomized controlled trials (RCTs) do not always engage with the intervention as its designers intended.
However, compliance in experimental education research is rarely a matter of yes-or-no, and participants who are exposed to an intervention but do not use it may still experience some effects. 
In general, researchers may be interested in estimating effects for subgroups of participants who implemented an intervention differently---that is, categorizing participants by \emph{how} they complied, rather than \emph{whether} they complied.

Estimating effects for compliance groups is tricky.
Consider, for example, an RCT testing an online learning platform that---among other features---allows students to request hints.
Could we estimate the effect of the intervention for the subset of students who requested hints? 
Causal inference in statistics typically depends on comparisons of outcomes between similar groups of subjects. 
However, there is no readily identifiable comparison group for hint users, who are only represented in the treatment arm, and who presumably struggled more frequently, on average, than their non-hint-user peers.

One approach to this quandary is based on principal stratification~\citep{frangakis}, a framework for defining and estimating effects for post-treatment subgroups, that is, subgroups that may themselves be a function of treatment assignment. It relies on the notion of~\emph{potential} subgroups called principal strata: which subgroup would a subject be in if they were (perhaps counterfactually) assigned to one treatment condition or another.
For example, the appropriate comparison group for hint-users is the set of students, assigned to the control arm, who~\emph{would have} requested hints if only given that option.
However, principal strata are unobserved---we cannot observe what choices students would have made had they been randomized differently. 
Nonetheless, researchers have devised estimators of principal effects---average effects within principal strata---to address a broad range of applied statistical problems.

\label{background}The conceptual basis for principal stratification---though not the terminology---emerged from the study of randomized experiments with noncompliance. Specifically,~\citet{air} showed that the econometric instrumental variables estimand is the average effect of the treatment on the principal stratum of ``compliers''---those subjects who would receive the treatment if and only if randomized to the treatment condition. Moreover, the average effect for compliers is nonparametrically identified if a random treatment assignment only has an effect for compliers---the ``exclusion restriction''---and if treatment assignment never~\emph{prevents} any subjects from receiving treatment (``monotonicity'').
In a follow-up analysis,~\citet{imbens1997bayesian} relaxed the exclusion restriction by way of parametric Bayesian finite mixture modeling, which remains a popular approach.

\citet{frangakis} introduced the terminology of principal stratification, and extended the framework of \citet{air} beyond questions of compliance to the study of post-treatment subgroups in general, inspiring a broad literature introducing (and debating) novel applications and estimators.
For instance,~\citet{fellerEtAl2016} estimated differential effects of random assignment to the Head Start early childhood care program for children who, if assigned to control, would receive home-based care, and for those who would attend a different early childhood care program.
Other methodologists have developed estimators for evaluating surrogate outcomes~\citep{li2010bayesian}, investigating causal mechanisms~\citep{lidsayPage}, addressing sample attrition~\citep{zhangRubin, ding2011}, and other related problems.
Since interest is often in comparing effects between principal strata---a goal that runs counter to the exclusion restriction---the earliest principal stratification estimators built on~\citet{imbens1997bayesian}'s parametric mixture models~\citep[e.g.][]{Barnard01062003,mealli2004analyzing}, and that approach remains popular~\citep{lidsayPage,fellerEtAl2016}.
However, these models are often difficult to specify and can be either hard or impossible to test.
Even when well-specified, they can yield severely biased and misleading estimates~\citep {griffin2008application,feller2016principal}.

Parametric mixture models are often difficult for applied researchers to implement---it is far removed from the regression models that dominate quantitative social science.
Fitting them may require familiarity with Bayesian Markov Chain Monte Carlo methods or likelihood optimization routines---even when these are automated~\citep[e.g.]{PStrata}, they are slow and require an additional round of diagnostics for model convergence.
Furthermore, standard model-checking techniques in regression analysis, such as examining residuals, are not straightforward to apply in these contexts.
Combined with their fragility and heavy dependence on model specification, these factors have severely limited their accessibility to applied researchers. 

These concerns have prompted the development of alternative approaches, including bounding~\citep{bounding} and randomization inference~\citep{nolen2011randomization}.
Of particular importance is the ``principal score,''~\citep{jo,dingLu,feller2017principal}, or the probability that a subject belongs to a particular principal stratum, as a function of pre-treatment covariates.
Under a strong, untestable assumption called ``principal ignorability,''  principal scores can be transformed into weights to compute an unbiased estimate of principal effects.
Principal ignorability is a direct analogue of the strong ignorability assumption in causal observational studies~\citep[e.g.][]{rosenbaum1983central}---that all covariates that predict both principal stratum membership and outcomes have been accounted for---in other words, the precise type of assumption that randomized trials are designed to avoid.
A separate set of results in~\citet{jo2002},~\citet{ding2011},
\citet{jiangDing2021},~\citet{richardson2023estimating}, and others have shown that, under certain circumstances, researchers can use baseline covariates to identify and estimate principal effects without relying on principal ignorability.

This paper extends those latter results by presenting a new principal effect estimator for one-way noncompliance, in which principal stratum membership is observed in one treatment group but not the other (future work will extend our approach to more general scenarios). Our estimator reduces to a simple three-step procedure: first, estimate principal scores, for instance, using logistic regression; then, impute the unknown principal subgroup membership indicators using the principal scores; and finally, estimate principal effects using ordinary least squares (OLS) regression with interactions. 
We show that this estimator is the solution to a set of estimating equations based on the combined moments of principal scores, indicators for treatment assignment, and observed principal stratum membership in the treatment group, outcomes, and covariates, and give a set of assumptions under which it is consistent for the true principal effects.

We will refer to the estimators presented here as ``\textsc{geepers}'': general estimating equations for principal effects using regressions.
\textsc{geepers} estimates rely on the correct specification of regression functions, but unlike common fully-parametric methods, they make no assumptions about the shape of the distribution of regression errors.
Unlike principal score weighting,~\textsc{geepers} does not assume principal ignorability and admits to sandwich-style standard error estimates~\citep{stefanskiBoos}, which we provide in the appendix.
Unlike any principal stratification estimator we are aware of,~\geepers~estimates are based on widely-used regression techniques, making them easy to compute, check, and understand for researchers without an extensive background in statistics. 
Existing methods fail to meet three needs: (1) accessibility, (2) robustness, and (3) model flexibility.~\textsc{geepers} addresses all three.

 
This work advances the literature in three ways: first, it makes an explicit connection between principal score methods, M-estimation, and semi-parametric estimation of principal effects using auxiliary data. Second, it provides a general estimator that relies on two of the most familiar tools in the applied education researcher's toolbox: linear and logistic regression. The \textsc{geepers} method can therefore make principal stratification methods accessible to a much broader range of applied education researchers. Lastly, this work includes an extensive simulation study that compares \textsc{geepers} to parametric mixture modeling and principal score weighting under a range of favorable and challenging scenarios. The simulation demonstrates the strengths and weaknesses of each method and the trade-offs between them.

The following section gives a formal introduction to principal stratification and briefly describes two common estimation techniques, parametric mixture modeling and principal score weighting.
Next, Section~\ref{sec:geepers} develops~\textsc{geepers} estimation and gives conditions for its consistency.
Section~\ref{sec:simulation} describes the simulation study. 
Section~\ref{sec:application} illustrates~\textsc{geepers} in a re-analysis of a set of RCTs conducted on an online middle-school math homework platform, designed to estimate the effects of being offered a particular type of assistance called video scaffolding. We will estimate effects for subgroups of students in the treatment arm who used the video scaffolds and those who did not. 
  Section~\ref{sec:discussion} concludes.

\section{Background}
In a randomized experiment with two conditions, let $Z_i=1$ if subject $i$, $i=1,\dots,n$ is assigned to the treatment condition, and let $Z_i=0$ otherwise.
If interest is in the effect of $Z$ on an outcome $Y$, under the ``stable unit treatment value assumption'' \citep{air} of no interference between units and no hidden versions of the treatment, define potential outcomes \citep{splawa1990application} $\yti$ as the value of $Y$ that $i$ would exhibit if $Z_i=1$, and let $\yci$ be the value if $Z_i=0$. Then the observed value of $Y$ satisfies $Y_i=Z_i\yti+(1-Z_i)\yci$, and the treatment effect of $Z$ on $Y$ for subject $i$ may be defined as $\tau_i=\yti-\yci$. The average treatment effect (ATE) is $\EE[\tau]$. %

Let $S_i\in\{0,1\}$ be a measurement on subject $i$ taken after treatment assignment, so that $S$ may be affected by $Z$.
Then $S$ itself has potential values $\sti$ and $\sci$, which $i$ would exhibit if $Z_i=1$ or $Z_i=0$, respectively.
Like potential outcomes $\yti$ and $\yci$, $\sci$ and $\sti$ are defined for all subjects, though $\sci$ is only observed when $Z_i=0$ and $\sti$ is only observed when $Z_i=1$. 
Assume ``strong monotonicity'' \citep[c.f.][]{dingLu} (also referred to as ``one-way noncompliance''):
\begin{ass}[Strong Monotonicity]\label{ass:sm}
\begin{equation*}
  \sci=0\text{ for all }i
\end{equation*}
\end{ass}
Under strong monotonicity, since $S$ is binary, every subject belongs to one of two ``principal strata,'' $\{\sci=0;\sti=1\}$ and $\{\sci=0;\sti=0\}$, or, more compactly, $\sti=1$ and $\sti=0$.
The average treatment effect in each principal stratum, $\eff1=\EE[\tau|\st=1]$ or $\eff0=\EE[\tau|\st=0]$, is called a principal effect. %

\label{page:PIcausal}Principal effects are causal---they describe the average effect of $Z$ on $Y$ for the subgroup of subjects with $\st=1$ or $\st=0$. 
However, the difference between the two principal effects, for instance, $\eff1-\eff0$, does not necessarily have a causal interpretation. 
For instance, in an RCT testing an intervention with two components, $S_i=1$ indicates that subject $i$ experienced both components, whereas subjects with $S=0$ experienced only the first.
If $\eff1-\eff0$ is positive, that may mean that exposure to the second component causes $Y$ to increase on average, or it may mean that the types of subjects who tended to be exposed to both components benefited more from the first component anyway. 

The challenge in estimating principal effects is that strata membership $\sti$ is only observed when $Z_i=1$, in which case $\sti=S_i$, the observed value of $S$.

\subsection{Estimating Principal Effects}
In estimating principal effects for a binary $S$, it will be useful to define four conditional means. For $s\in\{0,1\}$, let
\begin{equation}\label{eq:mus}
\mucs=\EE[Y_C|\st=s] \mbox{ and }\muts=\EE[Y_T|\st=s]
\end{equation}
Then, since
\begin{equation*}
  \effs=\muts-\mucs
\end{equation*}
estimating principal effects requires estimating $\muc0$, $\mut0$, $\muc1$, and $\mut1$.

We will focus on estimation in completely randomized experiments; that is, we will assume
\begin{ass}[Randomization]\label{ass:rand}
\begin{equation*}
  \yci,\yti,\sti, \bm{x}_i \independent Z_i
\end{equation*}
\end{ass}
where $\bm{x}_i$ is a vector of pre-treatment covariates, and $\independent$ denotes independence.

Under Assumptions~\ref{ass:sm} and~\ref{ass:rand}, conditional means $\mut0$ and $\mut1$ are nonparametrically identified.
In particular, the means of observed outcomes for subjects with $Z=1$ and $S=0$ or $S=1$ are unbiased for $\mut0$ and $\mut1$, respectively, since if $Z_i=1$, then $Y_i=\yti$ and $S_i=\sti$.
In contrast, $\muc0$ and $\muc1$ are not fully identified without further assumptions because $\sti$ and $\yci$ are never observed simultaneously (though they may be nonparametrically bounded; see, e.g., \citealt{bounding}).
We will briefly review two approaches to estimating $\muc0$ and $\muc1$
---normal mixture modeling and weighting---with some comments in-between on the role for covariates.

\subsubsection{Parametric Mixture Modeling}
Say an analyst assumes that, within principal strata, $Y_C$ is drawn from a parametric model, i.e., $f_{Y_C}(y|\st=s)=g^s(y)$. For instance, \citep{imbens1997bayesian} assumes $Y_C$ is normally distributed within strata, so $g^s(y)=\phi\left(\frac{y-\mucs}{\sigma_C^s}\right)$, where $\phi(\cdot)$ is the standard normal density function, and $\sigma_C^0>0$ and $\sigma_C^1>0$ are standard deviations. 
Then the probability density of $Y_C$ may be written as
\begin{equation}\label{eq:mixtureModel}
 f_{Y_C}(y)=Pr(\st=0)g^0(y)+Pr(\st=1)g^1(y).
\end{equation}
The probabilities $\Pr(\st=0)$ and $\Pr(\st=1)$ may be estimated first using data from the treatment group, since Assumption~\ref{ass:rand} implies that $\Pr(\st=1|Z=1)=\Pr(\st=1)$.
Given these estimates and~\eqref{eq:mixtureModel}, parameters $\muc0$ and $\muc1$ may be estimated using maximum likelihood, method of moments, or Bayesian techniques.

This approach has several problems.
First, the assumption of a conditional parametric model is untestable and limiting.
Unfortunately, even when it holds, mixture model estimates can be unreliable and, in some cases, severely biased \citep{griffin2008application,feller2017principal}.

\subsection{Principal Scores}
The principal score \citep[e.g.][]{jo} is defined as
\begin{equation}\label{eq:pscore}
  \ppi=\Pr(\sti=1|\bxsi),
\end{equation}
the probability of a subject with covariates $\bxsi$ being in the $\sti=1$ principal stratum.
Note that under randomization, $\Pr(\sti=1|\bxsi,Z_i)=\Pr(\sti=1|\bxsi)$, so a model for principal scores can be estimated using data from the treatment group and extrapolated to the control group.
Assume a model for principal scores:
\begin{equation}\label{eq:prinScoreMod}
  \ppip{\bm{\alpha}}=f(\bxsi;\bm{\alpha})
\end{equation}
with parameter vector $\bm{\alpha}$. An analyst may estimate $\bm{\alpha}$ as $\bm{\hat{\alpha}}$ by fitting model~\eqref{eq:prinScoreMod} using observed $\bm{S}$ and $\bxs$ for subjects with $Z=1$, and then compute estimated principal scores $\hppi=\ppip{\bm{\hat{\alpha}}}$ for subjects with $Z_i=0$.

Going forward, we will occasionally suppress dependence on $\bm{\alpha}$, and write $\ppip{\bm{\alpha}}=\ppi$.

If $\bm{x}$ is informative about $\st$, principal scores will vary between subjects. We will assume this is the case: 
\begin{ass}[Variable Principal Scores]\label{ass:vps}
 $\pp$ takes at least 3 distinct values
\end{ass}
For a broader discussion, see \citet{ding2011,jiangDing2021}.

Principal scores can potentially improve inference in a finite mixture model by modifying~\eqref{eq:mixtureModel} as
\begin{equation}\label{eq:mixtureModelPS}
  f_{Y_C}(y|\bm{x})=(1-\ppip{\bm{\alpha}})
g^0(y|\bm{x})+\ppip{\bm{\alpha}}
g^1(y|\bm{x})
\end{equation}
where $g^s(y|\bm{x})$, $s=0,1$, are conditional densities which must be specified. Most commonly, researchers assume a linear normal model, e.g., $g^s(y|\bm{x})=\phi\left(\frac{y-\mucs-\bm{\gamma}'\bm{x}}{\sigma_C^s}\right)$, where the columns of $\bm{x}$ are centered so that $\mathbb{E}[\bm{x}]=\bm{0}$. 

Alternatively, an analyst may assume ``principal ignorability'' \citep{jo,dingLu}:
\begin{equation}\label{eq:pi}
  Y_{C}\independent \st |\bxs
\end{equation}
or a somewhat weaker version, $\EE[Y_C|\st,\bxs]=\EE[Y_C|\bxs]$ \citep{feller2017principal}.
That is, the principal stratum is unrelated to control potential outcomes conditional on covariates.
Principal ignorability~\eqref{eq:pi} is reminiscent of ignorability assumptions typical in observational studies. 
In particular, an unobserved covariate $u$ that is correlated with both $\st$ and $Y_C$ would invalidate~\eqref{eq:pi}.

Under principal ignorability, $\muc0$ and $\muc1$ can be estimated via weighting:
\begin{equation}\label{eq:psw}
  \hat{\muc0}_w=\frac{\sum_{i: Z_i=0} Y_i(1-\ppi)}{\sum_{i:Z_i=0}1-\ppi}\text{ and } \hat{\muc1}_w=\frac{\sum_{i: Z_i=0} Y_i(\ppi)}{\sum_{i:Z_i=0}\ppi}
\end{equation}

Principal score weighting (\textsc{psw}) does not require any distributional assumptions and (after estimating principal scores) is very easy to implement.
On the other hand, the principal ignorability assumption is strong and restrictive.

\citet{feller2017principal} recommends the case-resampling bootstrap to estimate the sampling variances of $\hat{\mu}_{Cw}^0$, $\hat{\mu}_{Cw}^1$, and their associated principal effect estimators.

\subsection{M-Estimation}\label{sec:mest}
M-estimation (also known as the generalized method of moments or generalized estimating equations) is a general approach to deriving consistent statistical estimators and their asymptotic distributions.
In the following section, we will describe \textsc{geepers}, a novel principal effect estimator, and derive its properties using the principles of M-estimation.
Here, we provide a brief overview of M-estimation, primarily drawn from the excellent treatment in \citet{stefanskiBoos}.

Say observed data $\{Y_1,Z_1,S_1,\bm{x}_1\},\dots,\{Y_n,Z_n,S_n,\bm{x}_n\}\overset{\mathrm{iid}}{\sim} P$ for some distribution $P$, 
let $\bm{\theta}$ be a vector of parameters, possibly including principal score parameters $\bm{\alpha}$, potential outcome means $\bm{\mu}=\{\muc0,\muc1,\mut0,\mut1\}$, and coefficients from an outcome regression $\bm{\gamma}$, 
and let $\bm{\psi}(y,z,s,\bm{x};\bm{\theta})$ be a vector field such that $\mathbb{E}_P[\bm{\psi}(Y,Z,S,\bm{x};\bm{\theta}_0)]=\bm{0}$ for the true parameter value $\bm{\theta}_0$. 
Then, an M-estimator $\bm{\hat{\theta}}$ satisfies ``estimating equations'' $\sum_i \psi_j\left(Y_i,Z_i,S_i,\bm{x}_i;\bm{\hat{\theta}}\right)=0$. 
Some common examples are maximum likelihood estimators, where $\bm{\psi}$ is the score function 
or method-of-moments estimators for unidimensional data.

Under suitable regularity conditions, an M-estimator $\bm{\hat{\theta}}$ is asymptotically normal, and its sampling variance-covariance matrix may be estimated using the formula \citep[][ch. 7]{boosStefanskiBook}:
\begin{equation}\label{eq:sandwich}
  \widehat{var}(\bm{\hat{\theta}})=ABA'
\end{equation}
with matrices
\begin{equation*}
    A=\left\{\frac{1}{n}\displaystyle\sum_i\left. \frac{\partial \bm{\psi}(Y_i,Z_i,S_i,\bm{x}_i;\bm{\theta})'}{\partial \bm{\theta}'}\right|_{\bm{\theta}=\bm{\hat{\theta}}}\right\}^{-1}\mbox{ and }B=\frac{1}{n}\sum_i \bm{\psi}(Y_i,Z_i,S_i,\bm{x}_i;\bm{\hat\theta})\bm{\psi}(Y_i,Z_i,S_i,\bm{x}_i;\bm{\hat\theta})'.
\end{equation*}
Because of its form,~\eqref{eq:sandwich} is sometimes called a ``sandwich'' formula, where $A$ is the ``bread'' and $B$ is the ``meat'' \citep[e.g.][]{sandwich}. 
In general, the conditions for asymptotic normality can be difficult to establish; fortunately, they hold for OLS and for common generalized linear models such as logistic regression, provided the dimension of $\bm{\theta}$ does not increase with $n$.

Two factors recommend M-estimation for principal stratification: first, it requires only the correct specification of mean functions such as $\EE[Y|Z=z,\st=s,\bm{x}]$; other features of the population distribution, such as its shape or variance, do not need to be specified. 
Second, we will recommend a two-step procedure, wherein the analyst first fits the principal score model~\eqref{eq:prinScoreMod}, estimating $\bm{\alpha}$ as $\bm{\hat{\alpha}}$, and then uses $\bm{\hat{\alpha}}$ to estimate principal effects.
Two-step estimators fall neatly within the M-estimation framework, including sandwich variance estimators that propagate uncertainty from the first step into the second. 

\sloppy
Specifically, say that $\bm{\alpha}$ is estimated by solving estimating equations $\sum_i\bm{\psi^1}(Z_i,S_i,\bm{x}_i;\bm{\alpha})=\bm{0}$, and principal effects are estimated by solving a second set of estimating equations $\sum_i\bm{\psi^2}(Y_iS_i,Z_i,\bm{x}_i,\bm{\hat{\alpha}};\bm{\mu},\bm{\gamma})=\bm{0}$, where $\bm{\hat{\alpha}}$ is taken as a fixed input. 
The resulting estimator is the solution to the ``stacked'' estimating equations
\begin{equation}\label{eq:stacked}
 \sum_i \psi^{stack}(Y_i,Z_i,S_i,\bm{x}_i;\bm{\alpha},\bm{\mu},\bm{\gamma})=\displaystyle\sum_i\left\{\begin{array}{l} \psi^1(Z_i,S_i,\bm{x}_i;\bm{\alpha})\\\psi^2(Y_i,Z_i,S_i,\bm{x}_i;\bm{\alpha}, \bm{\mu},\bm{\gamma})\end{array}\right\}=\bm{0},
\end{equation}
A variance-covariance matrix estimated using the sandwich formula based on $\psi^{stack}$ accounts for uncertainty from both steps of the estimation procedure.

\section{GEEPERS }\label{sec:geepers}

The approach we introduce here incorporates principal scores into an M-Estimator for the mixture model~\eqref{eq:mixtureModel}.
Like principal score weighting, our M-estimator does not require distributional assumptions but instead requires a conditional independence assumption.
We will begin by describing a stronger-than-necessary conditional independence assumption to build intuition; in \S~\ref{sec:regression} we will present a considerably weaker alternative.

\subsection{Building Intuition: Covariate Ignorability}\label{sec:ci}

We begin by introducing an assumption that we term ``Covariate Independence'' (CI). It is a slightly weaker version of \citet{jiangDing2021}'s ``auxiliary independence'' assumption.
\begin{ass}[Covariate Independence]\label{ass:ci}
\begin{equation}\label{eq:assumption}
\EE[\yci|\bxsi,\sti]=\EE[\yci|\sti]=\muc0\text{ or }\muc1
\end{equation}
\end{ass}
i.e., $Y_C$ is mean-independent of $\bxs$ conditional on $\st$.
Under CI, covariates are not informative of the mean of $Y_C$ within principal strata.
As stated above, CI will rarely be plausible.
However, in some circumstances, researchers can identify a subset of observed covariates that satisfy CI and use those covariates in estimation.

It turns out that under strong monotonicity, randomization, and CI, and given a set of principal scores, the two principal effects $\eff0$ and $\eff1$ can be estimated by a simple OLS regression.
To see how we'll 
derive the estimating equations for $\muc0$, $\muc1$, $\mut0$, and $\mut1$, and show that they are equivalent to estimating equations for a model fit by OLS. 
The details of the calculations and the proofs are in the appendix.

The argument stems from a set of expressions for expectations, summarized in the following lemma:
\begin{lemma}\label{lemma:expectation}
Let \begin{equation}\label{eq:estEq0}
\tilde{\Psi}_i=\begin{pmatrix}
    (1-Z_i)\left[Y_i-\muc0-\ppi(\muc1-\muc0)\right]\\
    (1-Z_i)\left[\ppi Y_i-\ppi\muc0-\ppi^2(\muc1-\muc0)\right]\\
    Z_i\left[Y_i-\mut0-\sti (\mut1-\mut0)\right]\\
    Z_i\left[\sti Y_i -\sti\mut0-\sti^2(\mut1-\mut0)\right]
  \end{pmatrix}
\end{equation}
Then under Assumptions~\ref{ass:sm},~\ref{ass:rand},~\ref{ass:vps}, and~\ref{ass:ci}, $\EE[\Psi_i]=\bm{0}$.
\end{lemma}

Lemma~\ref{lemma:expectation} implies a set of estimating equations $\sum_{i=1}^n\tilde{\Psi}_i=\bm{0}$.

Now, the estimating equations $\sum_i\tilde{\Psi}_i=\bm{0}$ are algebraically equivalent to the estimating equations for a particular OLS fit, after some transformations. Most importantly, let
\begin{equation}\label{eq:ri}
\ri=Z_iS_i+(1-Z_i)\ppi=\begin{cases}
\ppi &Z_i=0\\
\sti &Z_i=1
\end{cases}
\end{equation}
In general, $(1-Z_i)\ppi=(1-Z_i)\ri$ and $Z_i\sti=Z_i\ri$, allowing $\ri$ to take the place of $\ppi$ and $S_i$ throughout $\tilde{\Psi}_i$.
Denote estimating equations based on $\ri$ as $\sum_i \psi_i=\bm{0}$.

\begin{prop}\label{prop:reg1}
  Under Assumptions~\ref{ass:sm},~\ref{ass:rand},~\ref{ass:vps},~\ref{ass:ci}, and suitible regularity conditions, if 
  $\ri$ is defined as in~\eqref{eq:ri} and the model
\begin{equation}\label{eq:regression0}
  \EE[Y_i|Z_i,\ri]=\beta_0+\beta_1\ri+\beta_2Z_i+\beta_3Z_i\ri
\end{equation}
is fit with OLS, yielding coefficient estimates $\hat{\beta}_k$, $k=0,\dots,3$, then
\begin{equation}
    \heff0_{CI}\equiv \hat{\beta}_2\mbox{ and }
    \heff1_{CI}\equiv \hat{\beta}_3+\hat{\beta}_2
\end{equation}
are M-estimators, with $\heff0_{CI}\rightarrow_p\eff0$ and $\heff1_{CI}\rightarrow_p\eff1$ as $n\rightarrow\infty$
\end{prop}

Thus, given principal scores, Assumption~\ref{ass:ci} enables an analyst to estimate principal effects with a simple regression.
Proposition~\ref{prop:reg1} builds on Theorem 2 of \citet{jiangDing2021}---which establishes identification of $\eff0$ and $\eff1$---by representing the principal effect estimators as a familiar regression using estimated principal scores.

If the principal scores are modeled as~\eqref{eq:prinScoreMod}, and parameters $\bm{\alpha}$ are estimated by M-estimation, with estimating equations $\sum_i Z_i\Omega_i=\bm{0}$, then the estimating equations for the principal score model and principal effect estimation can be stacked as in~\eqref{eq:stacked}.
Sandwich standard error estimates of the form~\eqref{eq:sandwich} based on $\left[Z_i\omega_i,\psi_i\right]$ will incorporate uncertainty from both principal score modeling and principal effect estimation conditional on principal scores.  

\subsection{Relaxing Covariate Independence with Outcome Modeling}\label{sec:regression}
In most applications, identifying a set of covariates $\bxs$ satisfying CI will be difficult or impossible.
Fortunately, CI may be relaxed or obviated by regression (see \citealt[][\S 3.4]{jiangDing2021} for an analogous identification result).

The assumption we state here is stronger than necessary but yields an attractively simple method; a weaker version of the assumption and the corresponding method, along with the proof, are provided in an appendix.

\begin{assp}{\ref*{ass:ci}$'$}[Residualized Covariate Independence]\label{ass:rci}
There exists $\bxy$, a (known) multivariate transformation of covariates $\bx$, and an (unknown) vector of coefficients $\bm{\gamma}$ such that
\begin{equation}\label{eq:ass:rci}
\begin{split}
\EE[\yc-\bm{\gamma}'\bxy|\bxs,\st]&=\EE[\yc-\bm{\gamma}'\bxy|\st]\\
\EE[\yt-\bm{\gamma}'\bxy|\bxs,\st]&=\EE[\yt-\bm{\gamma}'\bxy|\st]
\end{split}
\end{equation}
\end{assp}
That is, CI may not hold for potential outcomes themselves, but an analogous assumption holds for \emph{residualized} potential outcomes---i.e., $\yc$ and $\yt$ after subtracting out a linear function of (possibly transformed) covariates.
By forcing $\bm{\gamma}$ to be the same for both treatment groups, and (implicitly) for both principal strata, Assumption~\ref{ass:rci} assumes that there are no interactions between columns of $\bxy$ and either condition or principal stratum.
This is akin to the ``Additivity of Treatment Assignment Effect'' assumption in \citet{jo2002}.
Proposition~\ref{prop:interactions} in the appendix relaxes these requirements, but some preliminary simulations suggest that standard errors from estimates allowing for these interactions will be prohibitively large.
The simulation results in the following section address cases in which interactions are present in the data-generating model but not in the data analysis.

Assumption~\ref{ass:rci} allows principal effects to be estimated with an OLS model including $\bxy$:
\begin{prop}{\textsc{geepers}}\label{prop:reg2}

Under Assumptions~\ref{ass:sm},~\ref{ass:rand},~\ref{ass:vps}, and~\ref{ass:rci}, let principal scores be estimated as in~\eqref{eq:prinScoreMod} using data from the treatment group and assume they are linearly independent of $\bxy$. Then, say the researcher fits the following model with OLS:
\begin{equation}\label{eq:reg2}
Y_i=\beta_0+\beta_1\ri+\beta_2 Z_i+\beta_3Z_i\ri+\bm{\gamma}'\bxy_i+\epsilon_i
\end{equation}
where $\ri$ is defined as in~\eqref{eq:ri}.
Then let
\begin{equation}\label{eq:prinEffEst}
  \begin{split}
    \heff0_{RCI}&\equiv \hat{\beta}_2\\
    \heff1_{RCI}&\equiv \hat{\beta}_2+\hat{\beta}_3
  \end{split}
   \end{equation}
  $\heff0_{RCI}$ and $\heff1_{RCI}$ are M-estimators, with $\heff0_{RCI}\rightarrow_p\eff0$ and $\heff1_{RCI}\rightarrow_p\eff1$ as $n\rightarrow\infty$.
  Under suitable regularity conditions, they are jointly asymptotically normal, with a variance of the form~\eqref{eq:sandwich}.
\end{prop}
In short, researchers can estimate principal effects under strong monotonicity by imputing missing $\st$ values with estimated principal scores, fitting an OLS regression, and estimating standard errors using a sandwich estimator.
For the remainder of the paper, we will refer to principal effect estimators $\heff1_{RCI}$ and $\heff0_{RCI}$---our preferred estimators---as ``\textsc{geepers}.''

\section{A Simulation Study}\label{sec:simulation}

We conducted a simulation study to compare the performance of \textsc{geepers} with that of parametric mixture modeling (\pmm) and principal score weighting (\psw) estimators, in both favorable and unfavorable circumstances.
The study was designed to answer three sets of overarching questions.
First, how does the performance of the \textsc{geepers} estimator vary under different conditions, including sample size, the extent to which $\bx$ predicts $\st$, 
and when important interactions are omitted from the outcome model?
Second, how does \textsc{geepers} compare with parametric mixture modeling? Is it competitive in circumstances favorable to both techniques? Does it avoid the pitfalls of \pmm~when the parametric assumptions of the normal mixture model fail?
Third, are there ways or circumstances when \psw~outperforms \textsc{geepers} or \pmm~even though principal ignorability does not hold?

\subsection{Simulation Design}
The simulation study was conducted in \texttt{R} \citep{rcite} and Stan \citep{rstan}, and full replication code is available at [Redacted]. %
We replicated each set of conditions 5000 times.

\subsubsection{Data Generation}\label{sec:dataGeneration}

In each run of the simulation, we simulated three independent covariates, $x_k$, $k=1,\dots,3$, with $x_1,x_2\sim N(0,1)$, and the distribution of $x_3$ dependent on other factors but standardized so that $\EE[x_3]=0$, $Var(x_3)=1$.
Only the first two covariates were ``observed,'' i.e., were included in the analysis model.
Given all three covariates, the principal scores were set as
\begin{equation*}
  \ppf{}{x_1,x_2,x_3}=logit^{-1}\left[\alpha(x_{1i}-x_{2i}+x_{3i})\right]
\end{equation*}
where $\alpha$ is a manipulated factor.
Principal stratum $\sti\in \{0,1\}$ was simulated as $S_i\sim Bern(\ppi)$.

\label{potential}Potential and observed outcomes were generated as
\begin{equation}\label{eq:y-sim}
Y_i=0.3 Z_i\sti+(\gamma_1+\gamma_2\sti)(x_{1i}+x_{2i})+\gamma_3Z_ix_{1i}+\frac{1}{\sqrt{6}}x_{3i}+\epsilon_i
\end{equation}
with 
$\EE[\epsilon_i]=0$, and $Var(\epsilon_i)=1/2$.
The distribution of $\epsilon_i$ and the values of coefficients $\gamma_1$, $\gamma_2$, and $\gamma_3$ were manipulated factors in the study.

Treatment assignment $Z$ was simulated to be independent of all other variables, with half of the subjects assigned to treatment $Z=1$ and the rest to control $Z=0$.

Since the covariates were centered and independent of $Z$, the true principal effects were 
$\eff0=\gamma_3\EE[x1+x2|\st=0]$ and $\eff1=0.3+\gamma_3\EE[x1+x2|\st=1]$; in most simulation scenarios $\gamma_3=0$ so $\eff0=0$ and $\eff1=0.3$.

\subsubsection{Manipulated Factors}\label{sec:manipulatedFactors}
We manipulated four factors in the simulation, listed in Table~\ref{tab:factor}.
The factors were completely crossed, except that some levels of $n$ and $\alpha$ were tested with the other factors held fixed. 


\begin{table}[t]
    \caption{\label{tab:factor} Manipulated factors in simulation study}
  \centering
\begin{tabular}{r|ll}
  \hline
  Factor &\multicolumn{2}{l}{Levels}\\
  \hline
Sample size per condition $n$ &\multicolumn{2}{l}{$100^*,200^*,300^*,400^*,500,600^*,700^*,800^*,900^*,1000$}\\
$\alpha$ &\multicolumn{2}{l}{$\rd{0},0.2,0.3^\dagger,0.4^\dagger,0.5,0.6^\dagger,0.7^\dagger,0.8^\dagger,0.9^\dagger,1^\dagger$}\\
The distribution of $\epsilon_i$&\multicolumn{2}{l}{ Normal, \rd{Lognormal}
\rd{Uniform}}\\
\multirow{4}{*}{Interactions in~\eqref{eq:y-sim}}& None:  $\gamma_1=1/\sqrt{6};\gamma_2=\gamma_3=0$\\ 
& \rd{$\bm{x}:\st$}: $\gamma_1=3/(4\sqrt{6})$; $\gamma_2=1/(2\sqrt{6})$; $\gamma_3=0$\\
& \rd{$\bm{x}:Z$}: $\gamma_1=1/\sqrt{6};\gamma_2=0$; $\gamma_3=1/(2\sqrt{6})$\\
& \rd{$\bm{x}:\st$ \& $\bm{x}:Z$}: $\gamma_1=3/(4\sqrt{6})$; $\gamma_2=\gamma_3=1/(2\sqrt{6})$\\
\hline
\multicolumn{3}{p{4.5in}}{\footnotesize Factor levels in red violate assumptions of \geepers, \pmm, or both. Factors fully crossed except where indicated.}\\
\multicolumn{3}{p{4.5in}}{\footnotesize $*$ Only tested with $\alpha=0.5$, normal $\epsilon$, no interactions in~\eqref{eq:y-sim}.}\\
\multicolumn{3}{p{4.5in}}{\footnotesize $\dagger$ Only tested with $n=500$, normal $\epsilon$, no interactions in~\eqref{eq:y-sim}.}

\end{tabular}
\end{table}

\textbf{\emph{Sample Size per condition $\bm{n}$:}} The guarantees of M-estimation are all asymptotic. Hence, we examined the behavior of \textsc{geepers} at a range of sample sizes to determine its finite-sample properties. The range of 100--1000 was chosen to represent typical student-level randomized trials in education, for instance, \citet{growthMindsetRuralBurnette} with $n=115$ in the treatment group and $n=107$ in the control group, or \citet{impactPaper}, which initially randomized an average of 1000 students to each condition.

$\bm{\alpha}$\textbf{:} $\alpha$ controlled the extent to which $\st$ is predictable as a function of covariates. When $\alpha$ was higher, $\st$ was more easily predicted by covariates; when $\alpha=0$, Assumption~\ref{ass:vps} was violated, i.e., $Var(\pp)=0$. We expected that more predictive covariates would yield better performance for all three estimators.
To aid in the interpretation of the $\alpha$ parameter, the online appendix plots $\alpha$ against the area under the receiver operating characteristic curve for each fitted model (AUC), a common measure of classification accuracy \citep{bradley1997use}---an AUC of 0.5 implies that the model is no better than random guessing, and AUC of 1 implies perfect prediction. For $0.5\le \alpha \le 1$, the average AUC varies from roughly 0.53 to 0.77; for $\alpha=0.5$, the average AUC was roughly 0.675.
For reference, in the application in Section~\ref{sec:application}, the main model had an AUC of roughly 0.7. 

\textbf{\emph{The distribution of $\bm{\epsilon_i}$:}} $\epsilon$ is the regression error in~\eqref{eq:y-sim}. Across simulation runs, $E[\epsilon]=0$ and $Var(\epsilon)=1/2$. However, we varied the shape of $\epsilon$'s distribution between the normal distribution, as assumed by \pmm, the lognormal distribution, and the uniform distribution with minimum and maximum $\pm \sqrt{6}/2$. 
Since the covariate $x_3$ was unobserved and, therefore, essentially part of the regression error, we drew $x_3$ from the same distribution as $\epsilon$ and then standardized it to have a mean of 0 and a standard deviation of 1. 



\textbf{\emph{Interactions between $\st$ or $Z$ and covariates:}} This factor was designed to test the vulnerability of the estimators to certain violations of Assumption~\ref{ass:rci}. 

When neither interaction was present, $\gamma_2=\gamma_3=0$ and $\gamma_1=1/\sqrt{6}$.
In these cases, since  $Var(\gamma_1(x_1+x_2+x_3))=1/2$ and $Var(\epsilon_i)=1/2$, covariates explained half of the variance of $Y$ within treatment group and principal stratum, and observed covariates $x_1$ and $x_2$ explained 1/3 of that variance.
These values 
are typical in education contexts \citep{hedgesHedberg}.

When there was an interaction between covariates and $\sti$, then $\gamma_1=3/(4\sqrt{6})$ and $\gamma_2=1/(2\sqrt{6})$, so that the slopes for $x_1$ and $x_2$ varied by half their magnitude between principal strata, and on average they were equal to $1/\sqrt{6}$, as in the no-interaction case.
Interactions between covariate $x_1$ and $Z$ were controlled by $\gamma_3$: when there was an interaction, $\gamma_3=1/(2\sqrt{6})$, otherwise $\gamma_3=0$. 

Either type of interaction violates Assumption~\ref{ass:rci} since if $\gamma_2\ne 0$ or $\gamma_3 \ne 0$, there is no single vector $\bm{\gamma}$ that can fully capture the dependence of $\yc$ and $\yt$ on $S$---instead, the dependence varies with $\st$ and/or with $Z$.

\subsubsection{Analysis Models}\label{sec:simMods}

Using each simulated dataset, we estimated $\eff0$ and $\eff1$ with three methods: \geepers, \pmm~with principal scores~\eqref{eq:mixtureModelPS}, and the \psw~estimator~\eqref{eq:psw}. 

The methods all used the same ``observed'' dataset, consisting of outcomes $Y$, treatment assignments $Z$, two covariates $\bx=[x_1,x_2]$, and $S$ observed only when $Z=1$.
The third covariate used in the data-generating model, $x_3$, was ``unobserved''---since $x_3$ was correlated with both $\st$ and $Y$, Principal Ignorability~\eqref{eq:pi} was violated.

All three methods used the same principal score model, a logistic regression of $S$ on $\bx$ and an intercept, fit using data from the treatment group in which $S=\st$ was observed, i.e., $\ppf{\alpha_0,\bm{\alpha}}{\bx}=logit^{-1}(\alpha_0+\bm{\alpha}'\bx)$.
This model is very slightly misspecified: 
the principal scores, conditional on only $x_1$ and $x_2$, were $\ppf{}{x_1,x_2}=\Pr(\st=1|x_1,x_2)=\EE_{x_3}\left[\ppf{}{x_1,x_2,x_3}|x_3\right]$, which differs slightly from the logistic model.
This misspecification is distinct from the usual omitted variable bias---\geepers~and \pmm~do not assume that all predictors of $\st$ are included in the model---the issue is not the omission of $x_3$ \textit{per se}, but, rather, that logistic regression is not the correct model for $\Pr(\st|x_1,x_2)$. 

\pmm, the parametric mixture estimator, then fits two outcome models:
for the treatment group, the regression model $Y=\beta_0^T+\beta_1^TS+\bm{\gamma}'\bx+\epsilon$ and for the control group, a mixture model of the form~\eqref{eq:mixtureModelPS} with linear normal models for $g^s(y|x_1,x_2)=\phi\{(y-\mu^s_C-\bm{\gamma}'\bm{x})/\sigma_C\}$ in which the coefficients for covariates, $\bm{\gamma}$ were constrained to be equal across treatment groups and principal strata. 
Residual variance was allowed to differ across treatment groups, but not across principal strata; \label{page: priors}potential outcome means $\mucs$ and $\muts$, $s=0,1$ were given \label{page:prior}standard normal prior distributions, and other parameters were given reference priors. 
These models were fit simultaneously using Bayesian Markov Chain Monte Carlo with Stan \citep{rstan}. We interpreted the posterior mean as a point estimate and the posterior standard deviation as a standard error.

For \textsc{geepers} and \psw, the principal score model was fitted to data from the treatment group using standard maximum likelihood methods.
For \textsc{geepers}, we used the estimated principal scores in the control group, along with the observed $S$ in the treatment group, to construct $R$, and then fit the model~\eqref{eq:reg2} with OLS.
To estimate standard errors, we computed the sandwich covariance matrix following the procedure detailed in the appendix.

For \psw~we plugged estimated principal scores into~\eqref{eq:psw} to estimate $\muc0$ and $\muc1$, estimated $\mut0$ and $\mut1$ with the sample means of subjects with $Z=1$ and $S=0$ or $S=1$, respectively, and estimated $\heff0=\hat{\mu}_{Tw}^0-\hat{\mu}_{Cw}^0$ and $\heff1=\hat{\mu}_{Tw}^1-\hat{\mu}_{Cw}^1$.

\subsection{Results}\label{sec:simResults}
In this section, we evaluate estimates of $\eff0$. Results for estimates of $\eff1$, which are similar, can be found in the online appendix. 
In replication $b=1,\dots,5000$, define the error of a given estimator $\heff0_b$ as $err_b=\heff0_b-\eff0$. The bias of $\heff0$, $\EE[\heff0-\eff0]$, is estimated as $\widehat{bias}=\sum_b err_b/5000$, and the empirical standard error is defined as $\left\{\sum_b(\heff0_b-\overline{\heff0})^2/4999\right\}^{1/2}$, the sample standard deviation of $\heff0$. We also report the empirical coverage of confidence intervals, the proportion of replications in which $\eff0$ is within the estimated confidence bounds of $\heff0$. Finally, we estimated type-I error rates for level-0.05 tests of $H_0:\eff0=0$ for cases when $\gamma_3=0$ (i.e. no interactions between $x_1$ and $Z$ in the data generating model, in which cases $H_0$ is true) the proportion of the 5,000 runs in which $\heff0>2\widehat{SE(\heff0)}$, where $\widehat{SE(\heff0)}$ is the standard error estimated using the sandwich formula \eqref{eq:sandwich} for \geepers or the MCMC posterior standard deviation estimate for \pmm. Since the \psw estimator does not estimate standard errors, we only report on confidence interval coverage and type-I error rates for \geepers and \pmm. Results on root-mean-squared-error 
$\left(\sum_b err_b^2/5000\right)^{1/2}$ are in the online appendix.

\subsubsection{Performance with Varying $n$ and $\alpha$}
To what extent do the performances of \geepers~and \psw~depend on $n$ and $\alpha$?
For this set of results, $\epsilon\sim \mathcal{N}(0,1/2)$ and there were no interactions in~\eqref{eq:y-sim}; $\alpha$ was fixed at $0.5$ for varying levels of $n$, and $n$ was fixed at 500 for varying levels of $\alpha$. 

\begin{figure}[!ht]
  \centering
  \includegraphics[width=4.5in]{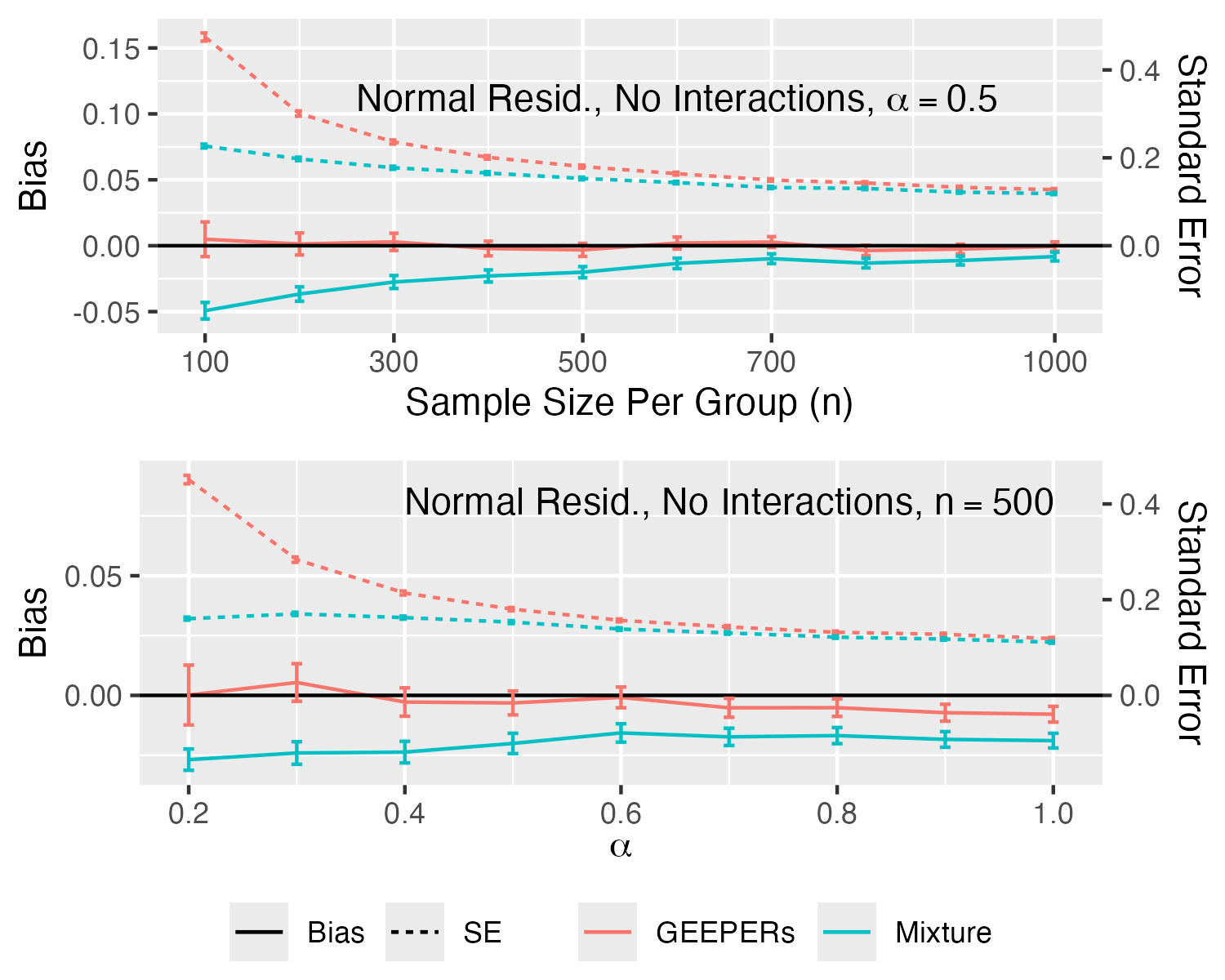}
  \caption{Empirical bias and standard error of \textsc{geepers} and Mixture estimates of $\eff0$ as sample size per treatment group $n$ or $\alpha$ varied. There were 5000 replications for each $n$ or $\alpha$. Other factors were held fixed at the noted values. Error bars show 95\% confidence intervals, representing simulation error.}
  \label{fig:alphan}
\end{figure}

Figure~\ref{fig:alphan} shows the empirical bias and standard error (SE) of the \textsc{geepers} and \pmm~estimators of $\eff0$ as sample size and $\alpha$ varied.
Across values of $n$, with $\alpha=0.5$, \textsc{geepers} was approximately unbiased, while \pmm~estimators exhibited slight negative bias that decreased with $n$. 
For values of $\alpha\le 0.6$, \geepers~appeared to be approximately unbiased, while the negative bias of the \pmm estimator decreased slightly with increasing $\alpha$. 
In contrast, both estimators exhibited slight negative bias for $\alpha>0.6$---possibly as a result of misspecification of the principal score model, as discussed above---with \geepers~remaining significantly less biased than \pmm. 

\geepers~had a higher SE than \pmm, particularly when $n=100$ or $\alpha=0.2$. The \pmm~advantage is far less dramatic for higher levels of $n$ and $\alpha$, eventually diminishing to approximately zero.

\subsubsection{Performance when Assumptions do not Hold}

Our next set of results compares the performance of \geepers, \pmm,
and \psw~when one or more assumptions of each method may be violated.

\begin{landscape}
\begin{figure}[!ht]
  \centering
  \includegraphics{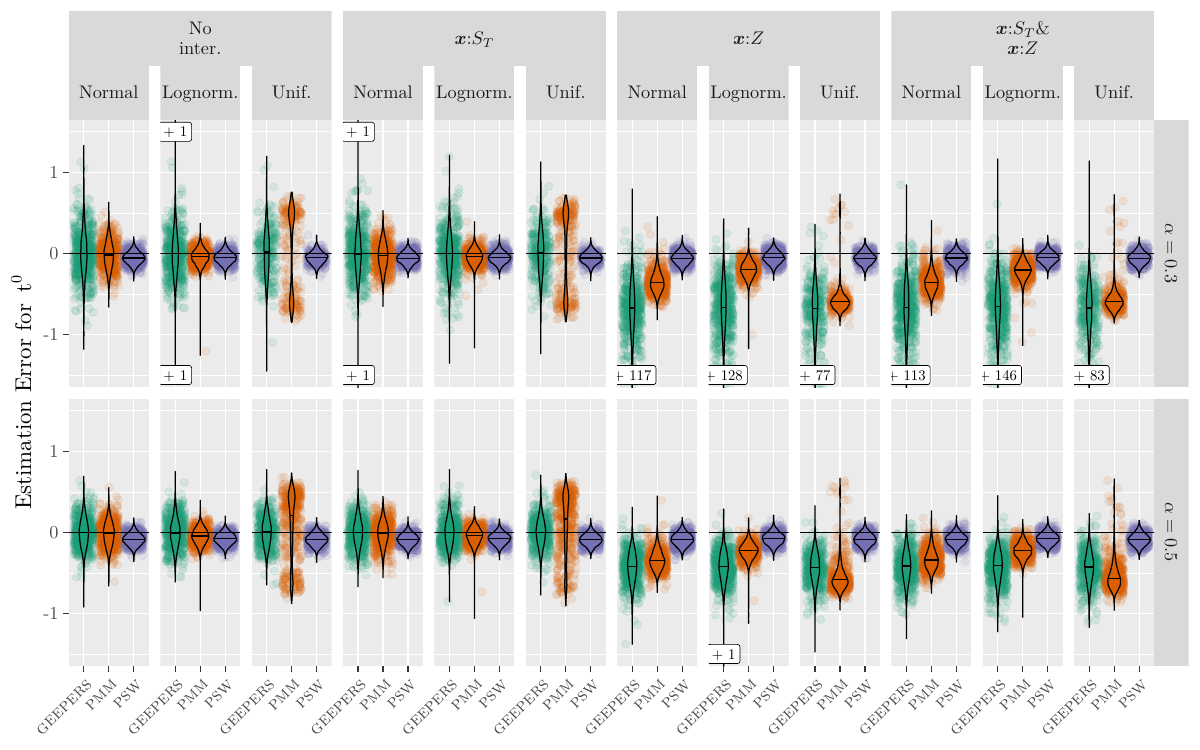}
  \caption{Violin and scatter plots of 5000 simulation estimation errors for estimators described in Section~\ref{sec:simMods} under varying conditions. For all plots, $n=500$; there were 5000 replications for each set of conditions. Annotations indicate the number of extreme outliers excluded from the plotting area.}
  \label{fig:boxplots}
\end{figure}

\end{landscape}

Figure~\ref{fig:boxplots} shows violin plots, layered on top of jittered scatter plots, showing estimation errors for \geepers, \pmm, and \psw. Tables~\ref{tab:coverage} and~\ref{tab:typeIerror} show empirical coverage rates of nominal 95\% confidence interval estimates of $\eff0$ and type-I error rates for level-0.05 hypothesis tests of $H_0: \eff0=0$ for \geepers~and \pmm; when at least one of an estimator's assumptions is violated, the result is colored red.

\begin{table}

\caption{\label{tab:coverage}Empirical coverage of nominal 95\% confidence intervals for estimates of $\eff0$.}
\centering

\begin{threeparttable}
\begin{tabular}[t]{lllllllll}
\toprule
\multicolumn{3}{c}{ } & \multicolumn{2}{c}{$\alpha=0$} & \multicolumn{2}{c}{$\alpha=0.3$} & \multicolumn{2}{c}{$\alpha=0.5$} \\
\cmidrule(l{3pt}r{3pt}){4-5} \cmidrule(l{3pt}r{3pt}){6-7} \cmidrule(l{3pt}r{3pt}){8-9}
\makecell[l]{Residual\\Dist.} & \makecell[l]{$\bm{x}:Z$\\Int.?} & \makecell[l]{$\bm{x}:S_T$\\Int.?} & \textsc{geepers} & \textsc{pmm} & \textsc{geepers} & \textsc{pmm} & \textsc{geepers} & \textsc{pmm}\\
\midrule
 & No & No & \rd{1.00} & 1.00 & 0.96 & 0.96 & 0.96 & 0.95\\

 & Yes & No & \rd{0.92} & \rd{0.89} & \rd{0.41} & \rd{0.37} & \rd{0.39} & \rd{0.33}\\

 & No & Yes & \rd{1.00} & \rd{1.00} & \rd{0.97} & \rd{0.96} & \rd{0.96} & \rd{0.95}\\

\multirow{-4}{*}{\raggedright\arraybackslash Normal} & Yes & Yes & \rd{0.92} & \rd{0.91} & \rd{0.42} & \rd{0.38} & \rd{0.41} & \rd{0.36}\\
\cmidrule{1-9}
 & No & No & \rd{1.00} & 0.99 & 0.96 & 0.99 & 0.95 & 0.98\\

 & Yes & No & \rd{0.92} & \rd{0.98} & \rd{0.41} & \rd{0.72} & \rd{0.40} & \rd{0.55}\\

 & No & Yes & \rd{1.00} & \rd{0.99} & \rd{0.96} & \rd{0.99} & \rd{0.96} & \rd{0.98}\\

\multirow{-4}{*}{\raggedright\arraybackslash Lognormal} & Yes & Yes & \rd{0.92} & \rd{0.98} & \rd{0.41} & \rd{0.72} & \rd{0.40} & \rd{0.57}\\
\cmidrule{1-9}
 & No & No & \rd{0.99} & \rd{0.25} & 0.94 & \rd{0.27} & 0.96 & \rd{0.47}\\

 & Yes & No & \rd{0.85} & \rd{0.12} & \rd{0.34} & \rd{0.05} & \rd{0.36} & \rd{0.11}\\

 & No & Yes & \rd{0.99} & \rd{0.42} & \rd{0.96} & \rd{0.38} & \rd{0.96} & \rd{0.54}\\

\multirow{-4}{*}{\raggedright\arraybackslash Uniform} & Yes & Yes & \rd{0.84} & \rd{0.21} & \rd{0.36} & \rd{0.08} & \rd{0.38} & \rd{0.15}\\
\bottomrule
\end{tabular}
\begin{tablenotes}[para]
\item \textit{Note: } 
\item \footnotesize Based on 5000 replications. $n=500$. Simulation standard error $\approx 1/3$ percentage point. Estimates colored \rd{red} indicate cases where the assumptions of the model are not met.
\end{tablenotes}
\end{threeparttable}

\end{table}

\begin{table}

\caption{\label{tab:typeIerror}Estimated type-I error probabilities for level-0.05 hypothesis tests of $H_0:\eff0=0$.}
\centering

\begin{threeparttable}
\begin{tabular}[t]{lllllllll}
\toprule
\multicolumn{3}{c}{ } & \multicolumn{2}{c}{$\alpha=0$} & \multicolumn{2}{c}{$\alpha=0.3$} & \multicolumn{2}{c}{$\alpha=0.5$} \\
\cmidrule(l{3pt}r{3pt}){4-5} \cmidrule(l{3pt}r{3pt}){6-7} \cmidrule(l{3pt}r{3pt}){8-9}
\makecell[l]{Residual\\Dist.} & \makecell[l]{$\bm{x}:Z$\\Int.?} & \makecell[l]{$\bm{x}:S_T$\\Int.?} & \textsc{geepers} & \textsc{pmm} & \textsc{geepers} & \textsc{pmm} & \textsc{geepers} & \textsc{pmm}\\
\midrule
 & No & No & \rd{0.00} & 0.00 & 0.04 & 0.04 & 0.04 & 0.05\\

 & Yes & No & \rd{0.09} & \rd{0.12} & \rd{0.63} & \rd{0.68} & \rd{0.69} & \rd{0.75}\\

 & No & Yes & \rd{0.00} & \rd{0.00} & \rd{0.03} & \rd{0.04} & \rd{0.04} & \rd{0.05}\\

\multirow{-4}{*}{\raggedright\arraybackslash Normal} & Yes & Yes & \rd{0.08} & \rd{0.10} & \rd{0.62} & \rd{0.67} & \rd{0.67} & \rd{0.72}\\
\cmidrule{1-9}
 & No & No & \rd{0.00} & 0.01 & 0.04 & 0.01 & 0.05 & 0.02\\

 & Yes & No & \rd{0.08} & \rd{0.03} & \rd{0.63} & \rd{0.37} & \rd{0.70} & \rd{0.61}\\

 & No & Yes & \rd{0.00} & \rd{0.01} & \rd{0.04} & \rd{0.01} & \rd{0.04} & \rd{0.02}\\

\multirow{-4}{*}{\raggedright\arraybackslash Lognormal} & Yes & Yes & \rd{0.08} & \rd{0.02} & \rd{0.63} & \rd{0.37} & \rd{0.69} & \rd{0.60}\\
\cmidrule{1-9}
 & No & No & \rd{0.01} & \rd{0.75} & 0.06 & \rd{0.73} & 0.04 & \rd{0.53}\\

 & Yes & No & \rd{0.15} & \rd{0.88} & \rd{0.70} & \rd{0.95} & \rd{0.72} & \rd{0.91}\\

 & No & Yes & \rd{0.01} & \rd{0.58} & \rd{0.04} & \rd{0.62} & \rd{0.04} & \rd{0.46}\\

\multirow{-4}{*}{\raggedright\arraybackslash Uniform} & Yes & Yes & \rd{0.16} & \rd{0.79} & \rd{0.68} & \rd{0.92} & \rd{0.71} & \rd{0.88}\\
\bottomrule
\end{tabular}
\begin{tablenotes}[para]
\item \textit{Note: } 
\item \footnotesize Based on 5000 replications. $n=500$. Simulation standard error $\approx 1/3$ percentage point. Estimates colored \rd{red} indicate cases where the assumptions of the model are not met.
\end{tablenotes}
\end{threeparttable}

\end{table}

When all of the \geepers~and \pmm~assumptions were met, that is, when $\epsilon\sim\mathcal{N}(0,1/2)$, $\alpha>0$, and there were no covariate interactions in~\eqref{eq:y-sim}, results for $\geepers$ and $\pmm$ were the same as in the previous section; in addition, confidence intervals from both methods achieved their nominal levels. 
\textsc{psw} had by far the lowest variance and the highest bias of the three estimators; the latter is due to the violation of principal ignorability stemming from the omission of $x_3$ in the analysis. 

When $\alpha=0$, Assumption~\ref{ass:vps} is violated; 
nevertheless, as the first column of results in Table~\ref{tab:coverage} shows, when other assumptions were met, confidence intervals from both \geepers~and \psw~tended to over-cover when $\alpha=0$. 
In a randomized experiment, $\alpha$ can be estimated directly, making Assumption~\ref{ass:vps} testable.

The \pmm estimator assumes that outcomes are normally-distributed conditional on covariates, treatment assignment, and principal stratum, while \geepers and \psw do not.
Our results suggest that \pmm is sensitive to some departures from normality but insensitive to others---specifically, \pmm sampling distributions and coverage and type-I error rates when errors were drawn from a lognormal distribution resemble those from when errors are normally distributed. On the other hand, when errors are drawn from a uniform distribution, \pmm sampling distributions appeared bimodal, confidence intervals severely undercovered, and type-I error rates far exceeded 0.05. 
In contrast, the performance of \geepers and \psw largely does not depend on the distribution of the regression errors.

The simulation design explored two types of misspecification of the outcome model: the presence of interactions between covariates $x_1$ and $x_2$ and principal stratum $\st$ (i.e. $\gamma_2=1/(2\sqrt{6})$) or between $x_1$ and treatment assignment $Z$. 
Interactions between $\bm{x}$ and $\st$ had little or no impact on the performance of any of the estimators.
On the other hand, interactions between $x_1$ and $\st$ led to a large negative bias in \geepers and \pmm estimates, resulting in confidence interval undercoverage and inflated type-I error rates. 

This result is disappointing---interactions between covariates and treatment assignment within principal strata can be hard to identify and diagnose---but not fatal. 
The consistency of the \geepers~estimator does not depend on the inclusion of all relevant covariates, so if researchers suspect treatment effects to vary with a particular covariate, they may refit the \geepers estimator excluding that covariate.
To illustrate, we refit all three estimators to the datasets in which $Z:x_1$ interactions were present; Table~\ref{tab:nox1} compares bias and coverage between analyses including and excluding $x1$ (for simplicity, Table~\ref{tab:nox1} only includes results from cases without $\bm{x}:\st$ interactions). 
When $x_1$ is excluded from the analysis, the biases for \geepers return to near-0 and associated confidence intervals achieve their nominal levels; the same holds for \pmm when regression errors are normally or lognormally distributed.
In contrast, the \psw bias increases by roughly a factor of 2, since excluding $x_1$ exacerbates principal ignorability violations. 
These results suggest a method of sensitivity analysis of refitting \geepers estimates excluding potential effect moderators; drastic differences in effect estimates may indicate problems. 

\begin{table}

\caption{\label{tab:nox1}Estimated bias and 95\% CI coverage rates when interactions between $x_1$ and $Z$ are present and $x_1$ is either excluded (``just $x_2$'') or included (``$x_1$ \& $x_2$'') in the analysis.}
\centering

\begin{threeparttable}
\begin{tabular}[t]{lrlrrrrrr}
\toprule
\multicolumn{3}{c}{ } & \multicolumn{6}{c}{Error Distribution} \\
\cmidrule(l{3pt}r{3pt}){4-9}
\multicolumn{3}{c}{ } & \multicolumn{2}{c}{Normal} & \multicolumn{2}{c}{Lognormal} & \multicolumn{2}{c}{Uniform} \\
\cmidrule(l{3pt}r{3pt}){4-5} \cmidrule(l{3pt}r{3pt}){6-7} \cmidrule(l{3pt}r{3pt}){8-9}
  & $\alpha$ & estimator & Just $x_2$ & $x_1$ \& $x_2$ & Just $x_2$ & $x_1$ \& $x_2$ & Just $x_2$ & $x_1$ \& $x_2$\\
\midrule
 &  & \textsc{geepers} & 0.06 & -0.69 & 0.09 & -0.69 & 0.05 & -0.70\\

 &  & \textsc{pmm} & -0.07 & -0.35 & -0.09 & -0.21 & -0.03 & -0.42\\

 & \multirow{-3}{*}{\raggedleft\arraybackslash 0.3} & \textsc{psw} & -0.12 & -0.06 & -0.11 & -0.05 & -0.12 & -0.06\\

 &  & \textsc{geepers} & 0.05 & -0.43 & 0.05 & -0.42 & 0.05 & -0.43\\

 &  & \textsc{pmm} & -0.06 & -0.34 & -0.11 & -0.23 & 0.02 & -0.49\\

\multirow{-6}{*}{\raggedright\arraybackslash Bias} & \multirow{-3}{*}{\raggedleft\arraybackslash 0.5} & \textsc{psw} & -0.17 & -0.09 & -0.17 & -0.08 & -0.18 & -0.09\\

\cmidrule{1-9}
 &  & \textsc{geepers} & 0.98 & 0.41 & 0.97 & 0.41 & 0.97 & 0.34\\

 & \multirow{-2}{*}{\raggedleft\arraybackslash 0.3} & \textsc{pmm} & 0.98 & 0.37 & 0.99 & 0.72 & 0.81 & 0.05\\

 &  & \textsc{geepers} & 0.96 & 0.39 & 0.96 & 0.40 & 0.96 & 0.36\\

\multirow{-4}{*}{\raggedright\arraybackslash Coverage} & \multirow{-2}{*}{\raggedleft\arraybackslash 0.5} & \textsc{pmm} & 0.97 & 0.33 & 0.98 & 0.55 & 0.76 & 0.11\\
\bottomrule
\end{tabular}
\begin{tablenotes}[para]
\item \textit{Note: } 
\item \footnotesize Based on 5000 replications. $n=500$. Simulation standard error $\approx 1/3$ percentage point. Estimates colored \rd{red} indicate cases where the assumptions of the model are not met.
\end{tablenotes}
\end{threeparttable}

\end{table}

\label{simsum}In sum, \psw~is the least variable of all three estimators, and does not depend on outcome modeling, but is vulnerable to bias from an unmeasured covariate that predicts both $S$ and $Y$.
When all assumptions of both methods are met, \geepers~is less efficient, but also less biased, than \pmm, but differences in efficiency decrease with $n$ and $\alpha$.
Neither \psw~nor \pmm~is vulnerable to interactions between $X$ and $S$.
\pmm~is highly vulnerable to certain violations of normality but not to others, while \geepers is not. 
Interactions between covariates and treatment assignment (i.e., effect moderators) pose a serious threat to both \geepers and \pmm, though sensitivity analyses excluding potential moderators may help.

\section{Application: an Educational A/B Test of Video Scaffolding}\label{sec:application}

We illustrate \geepers in a secondary analysis of a set of RCTs conducted within ASSISTments, an online homework platform for middle-school mathematics, using replication data from \citet{commonTrends}, available at \url{https://osf.io/59shv/}.

 Students work on math problems in ASSISTments, and on most problems receive immediate correctness feedback, along with the option to request hints or explanations. 
At the same time, ASSISTments supports embedded field experiments or A/B tests called ``E-Trials.'' Researchers propose contrasting conditions (e.g., alternative sets of hints) for specific modules or problems; when teachers assign these modules, students are individually randomized to conditions.

For this illustration, we re-analyze data combined from three experiments evaluating video scaffolding within ASSISTments Skill Builders, mastery-based modules in which students continue working problems until demonstrating mastery by correctly answering three problems in  a row without assistance. Scaffolding is triggered when a student errs, or requests help: the problem is decomposed into simpler sub-steps, and students input intermediate results \citep[see, e.g.,][]{mcguire}.
In the video-scaffolding condition, prompts and explanations are delivered via short videos. The control condition (``answer-only'') allows struggling students to receive a wrong mark, view the correct answer, and proceed.

These experiments were first analyzed, along with 47 others testing different interventions, in \citet{commonTrends}, which gathered, published, and analyzed data from 50 different E-trials experiments. 
Not all students randomized to video scaffolding actually viewed any scaffolds. 
Noncompliance could be for several reasons: some never required help, some exited the assignment rather than request help, some may have used alternative help sources, and some could not view videos.
Still, randomization to the treatment condition could conceivably influence outcomes even for students who never request scaffolds (e.g., via the offer of scaffolding, frustration among students unable to view videos, or other unrecorded differences between arms).
That said, if there is an effect of randomization for never-takers, it is likely to be relatively small. 

We estimated two principal effects: the average effects of randomization for students who would versus would not receive scaffolding under treatment.

\subsection{Data}
\begin{table}
\caption{Descriptive statistics for video scaffolding experiments.}
\label{tab:descriptives}
\centering

\begin{tabular}[t]{lrrrrrr}
\toprule
\multicolumn{1}{c}{ } & \multicolumn{2}{c}{Sample Size} & \multicolumn{2}{c}{Compliance} & \multicolumn{2}{c}{Mastery Speed} \\
\cmidrule(l{3pt}r{3pt}){2-3} \cmidrule(l{3pt}r{3pt}){4-5} \cmidrule(l{3pt}r{3pt}){6-7}
Experiment & Trt & Ctl & $\bar{S}|Z=1$ & Avg. \# Scaffolds|S=1 & Trt & Ctl\\
\midrule
A & 361 & 337 & 0.85 & 6.25 & 0.19 & -0.13\\
B & 27 & 45 & 0.85 & 7.96 & -0.25 & -0.05\\
C & 138 & 125 & 0.65 & 6.06 & 0.00 & -0.28\\
Total & 526 & 507 & 0.80 & 6.30 & 0.12 & -0.16\\
\bottomrule
\end{tabular}

\end{table}

The analytic sample includes 1,033 students (526 video-scaffolding; 507 answer-only). Table \ref{tab:descriptives} reports experiment-specific sample sizes.

For this illustration, we consider students in the treatment group to have complied if they received any video scaffolding assistance---i.e., let $S_i=1$ if a student $i$ received scaffolding on at least one problem, and $S_i=0$ otherwise ($S=0$ for all students randomized to control). Table \ref{tab:descriptives} reports the proportions of students in treatment groups of each experiment with $S=1$---overall, it was 80\%---and, among those who received any scaffolds, the number of problems on which they received scaffolding assistance. 

Table \ref{tab:descriptives} also reports average outcomes: the three video-scaffolding experiments all used  $n^{prob}_i$, the number of problems worked by student $i$, 
as their primary outcome. 
Since students were supposed to work until they demonstrated mastery, lower values of $n^{prob}$ should indicate faster learning. 
In the replication data for \citet{commonTrends}, $n_{prob}$ was tranformed into ``mastery speed,'' 
$$MS_i=\frac{-log(n^{prob}_i)-\hat\mu_{ex[i]}}{\hat\sigma_{ex[i]}}$$
where $\hat\mu_{ex[i]}$ and $\hat\sigma_{ex[i]}$ are the sample mean and standard deviation of $-log(n^{prob})$ for students in the same experiment as $i$. 

Besides $MS$ and randomization indicators $Z$, the dataset includes detailed student log data gathered during the experiment, and a set of student- and class-level covariates derived from students' prior ASSISTments log data. 
For a full set of available covariates and their descriptions, please see \citet{commonTrends}.

\subsection{Analysis}

\geepers requires specifying a model for principal scores and another model for outcomes as a function of treatment assignment, principal stratum, and covariates.
We conducted five such analyses, each using a different set of covariates. 
Within each analysis, both models used the same set of covariates---the principal score model using logistic regression and the outcome model using OLS.

Three criteria governed our choice of covariates and regression specifications. First, Proposition \ref{prop:reg2} requires consistent estimation of principal scores and (via Assumption \ref{ass:rci}) correct specification of the outcome model, that is, the
model for $\EE[Y|\bm{x},Z,S_T]$. 
Hence, it is essential to choose covariates and model specifications that lead to a good model fit.
Second, Assumption \ref{ass:vps} requires principal scores to vary with at least some element of $\bm{x}$. Moreover,
the results of the simulation study suggest that when $S_T$ can be predicted more precisely as a
function of $\bx$, \textsc{geepers} principal effect estimators are more precise and more robust to model
misspecification. 
Those considerations militate in favor of choosing covariates $\bx$ to optimize out-of-sample principal score prediction accuracy.
We also examined binned residual plots \citep{arm} for candidate principal score models and residual plots for outcome models to ensure adequate model fit. 

The first principal score model, labeled ``Full,'' included all available covariates. 
To protect against overfitting, the second model, labeled ``AIC Opt,'' included covariates chosen with a backward stepwise selection algorithm, minimizing the
AIC \citep{aic} of the principal score model. 
The AIC Opt. model included experiment indicators, the percent of students' prior skill builders that were completed, the average prior skill-builder completion for students classrooms, the number of prior problem sets each student and classroom had begun, the average number of attempts per problem each student had made before the experiment, class size, and prior correctness---the percent of prior problems each student answered corrrectly. 

Lastly, Assumption 4 implies that there is no interaction between $\bm{x}$ and $Z$ in the outcome
model. A priori, three student attributes pose the greatest threat to this assumption: prior achievement, prior skill builder completion, and indicators for the three experiments. 
Therefore, we estimated principal score models dropping covariates related to each of those constructs from the ``AIC Opt'' model: a model labeled ``-Pretest'' without student prior correctness, a model labeled ``-Completion'' dropping student- and class-level prior skill builder completion, and a model labeled ``-Exp. IDs,'' dropping experiment indicators. 

\section{Results}
\begin{table}
\caption{\label{tab:prinEffs}Estimated principal effects for the video scaffolding experiments using different analyses. Standard errors are in parentheses. AUCs from the principal score models were estimated with 10-fold cross-validation.}
\centering

\begin{tabular}[t]{ccccc}
\toprule
\multicolumn{3}{c}{ } & \multicolumn{2}{c}{Principal Effects} \\
\cmidrule(l{3pt}r{3pt}){4-5}
Method & PS Model & AUC & Never Takers & Compliers\\
\midrule
\textsc{geepers} & Full & 0.67 & -0.02 (0.36) & 0.37 (0.11)\\

 & AIC Opt & 0.71 & -0.04 (0.39) & 0.37 (0.12)\\

 & -Pretest & 0.70 & 0.21 (0.42) & 0.30 (0.12)\\

 & -Exp. IDs & 0.67 & -0.62 (0.52) & 0.50 (0.14)\\

 & -Completion & 0.71 & -0.54 (0.41) & 0.47 (0.12)\\
\cmidrule{1-5}
\textsc{psw} & AIC Opt &  & 0.80 (0.07) & 0.15 (0.07)\\

\textsc{pmm} &  &  & 0.35 (0.11) & 0.26 (0.07)\\

\textsc{iv} &  &  & 0.00 (0.00) & 0.32 (0.08)\\
\cmidrule{1-5} 
 \textsc{ATE} & & & \multicolumn{2}{c}{0.28 (0.06)}\\
\bottomrule
\end{tabular}
\end{table}

\begin{figure}[!ht]
  \centering
  \includegraphics[width=4.5in]{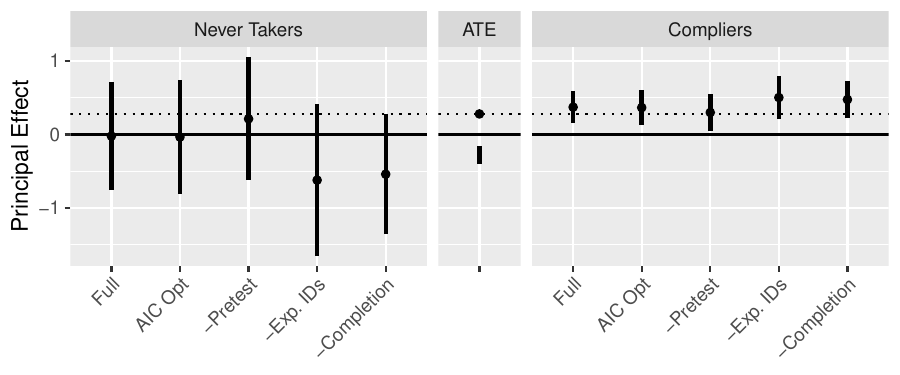}
  \caption{\geepers principal effect estimates for compliers and never-takers from the video scaffolding experiments, using five different models, along with a difference-in-means estimate for the ATE. The five principal stratification analyses differed in which covariates they used: ``Full'' used all available covariates, ``AIC Opt'' selected covariates using an AIC step-down procedure applied to the principal score model, and ``-Pretest,'' ``-ExpIDs,'' and ``-Completeness'' dropped prior achievement, experiment indicators, and prior skill builder completeness measures, respectively, from ``AIC Opt'' model. }
  \label{fig:prinEffs}
\end{figure}

The three principal score models were fairly successful in distinguishing students who use scaffolding---compliers---from those who do not---never takers. The AUC, evaluated using out-of-sample predictions in a 10-fold cross-validation, was about 0.67 for the Full and noExp models and around 0.70 for the other three models. That is, considering all possible pairs of students in which one has $S=1$, and the other $S=0$, the student with $S=1$ will have a higher principal score in roughly 67–70\% of cases. Coefficient estimates from the principal score models and from the outcome models are displayed in Tables \ref{tab:psMods} and \ref{tab:outModAppendix} in the online appendix. 
Diagnostic residual plots for principal score and outcome models from the five analyses (Figures \ref{fig:binnedResids} and \ref{fig:outResids} in the appendix) showed no indication of model misspecification. 

Figure \ref{fig:prinEffs} and Table \ref{tab:prinEffs} show principal effect point estimates and 95\% confidence intervals from analyses based on the five principal score models, along with a difference-in-means estimate of the ATE.
The ATE was positive: on average, randomization to the video scaffolding condition improved mastery speed by 0.28$\pm$0.12 standard deviations (SDs).
Principal effect estimates for compliers were likewise positive across all five analyses: analyses using the Full and AIC Opt models estimated principal effects of roughly 0.37$\pm$0.22 SDs, and analyses excluding potential effect moderators yielded similar estimates, ranging from 0.30$\pm$0.25 SDs for -Pretest to 0.50$\pm$0.29 SDs for -Exp. IDs, with confidence intervals largely overlapping. 

Since roughly 80\% of the students in the sample were compliers, estimated principal effects for never takers had much higher standard errors than for compliers. 
Analyses based on the Full and AIC Opt models yielded estimates of -0.02$\pm$0.73 and -0.04$\pm$0.7, respectively.
The -Pretest analysis estimated a positive principal effect, 0.21$\pm$0.84 and the -Exp IDs and -Completion analyses yielded negative principal effect estimates, 0.62$\pm$1.03 and -0.54$\pm$0.82, respectively. 
Due to their large standard errors, all five of these results are consistent with each other and with the absence of any effect for never takers. 

\section{Comparing Principal Stratification Methods}

\begin{figure}[!ht]
\centering
\includegraphics[width=4.5in]{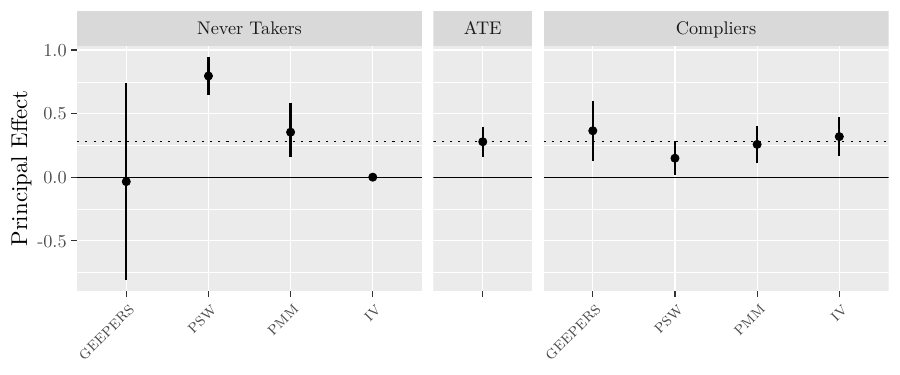}
\caption{Principal effect estimates for compliers and never-takers from the video scaffolding experiments, using four different methods---\geepers, parametric mixture modeling (\pmm), principal score weighting (\psw) and instrumental variables using two-stage-least squares regression (\textsc{iv})---along with a difference-in-means estimate for the ATE.}
\label{fig:compareMethods}
\end{figure}

Figure~\ref{fig:compareMethods} 
shows principal effect estimates from \geepers compared to analogous estimates using \psw, \pmm, and an instrumental variables estimator, alongside estimates of the overall average treatment effects, labeled ``ATE.''
All four methods used the covariate specification from ``AIC Opt.''

\psw estimated a large, positive effect for never-takers, 0.80$\pm$0.15 SDs, and a relatively small effect of 0.15$\pm$0.13 SDs for compliers. 
This result may, in part, be due to the failure of principal ignorability \eqref{eq:pi}: students receive scaffolding support if they are unsure how to solve a problem or if they make an error. 
Even conditional on covariates, we might expect students who requested support to have struggled more on the problem set than students who did not request support---in other words, 
$\EE[\yc|\st=1,\bx]<\EE[\yc|\st=0,\bx]$.
This, in turn, would induce positive bias for never takers and negative bias for compliers, consistent with the observed result.

The Bayesian \pmm assumed that outcomes were normally distributed, conditional on
covariates and principal strata, with a residual standard deviation that varied between principal
strata. 
All model parameters with support in $\mathbb{R}$ were given standard normal priors, while standard
deviation parameters were given half-standard normal priors.
\pmm estimated a similar effect as \geepers for compliers, albeit with a smaller standard error--0.26$\pm$0.14 SDs. 
For never takers, \pmm estimated a relatively large positive effect of 0.35$\pm$0.19. 
This latter result is surprising, and suspicious---it seems unlikely that the effect of randomization would be larger for never takers than for compliers, and a graphical posterior predictive check (Figure \ref{fig:postPred} in the appendix) uncovers marked misspecification of the outcome distribution. 

Our final comparison is with a standard instrumental variables estimator, as implemented in the \texttt{AER} package in \texttt{R} \citep{AER}.
Unlike \geepers, \psw, and \pmm, instrumental variables estimators assume the ``exclusion restriction'' of no effect of randomization on never takers' outcomes, and only estimate effects for compliers. 
As described above, in the context of the video scaffolding experiments, this is a reasonable assumption, but it is not guaranteed. 
The instrumental variables estimate in \ref{fig:compareMethods} used linear regression adjustment for the same covariates as in the AIC Opt. model, and used a heteroskedasticity-consistent standard error estimator from the \texttt{sandwich} package in \texttt{R} \citet{sandwich}. 
The estimated average effect for compliers, 0.32$\pm$0.15 SDs, is quite similar to the \geepers estimate, albeit with a smaller standard error.

\section{Discussion}\label{sec:discussion}
\textsc{geepers} is a straightforward approach to principal effect estimation under strong monotonicity, which is more robust---though sometimes less precise---than alternative approaches under a wide array of scenarios. It is built on widely used regression models, making it accessible to applied researchers. 

Like all other principal stratification methods, it is sensitive to violations of its underlying assumptions---in particular, the simulation results showed that including treatment effect moderators in the principal score model can lead to bias. 
Fortunately, \geepers does not assume that all relevant covariates are included, so researchers may conduct sensitivity checks by excluding suspected moderators from their analyses. 
However, dropping important predictors of principal stratum membership can also increase standard errors. 
One promising solution, which is the subject of ongoing research, is to incorporate automatic regularization into either principal score or outcome models. 
Proper regularization of the outcome regression can alleviate collinearity between principal scores and covariate regressors, which, in turn, may allow for a weakening of assumption \ref{ass:rci}.
Future research will explore avenues for automatic regularization, including methods for accounting for regularization in standard error estimates, and determine when regularization can produce approximately unbiased estimates. 

There is good reason to hope that extensions to \textsc{geepers}, including cases in which $S$ takes more than two values, may be straightforward.
For instance, in truncation by death problems \citep[e.g.][]{zhangRubin,ding2011} with weak monotonicity, $S_i=1$ if participant $i$ survives (or, more generally, if the outcome $Y_i$ is measured) and 0 otherwise, interest is typically in the principal effect for the ``always survive'' principal stratum in which $\sti=S_{ci}=1$.
If, say, $\sti\ge S_{Ci}$, then every subject in the control condition with $S_i=1$ is in the always survive stratum, while those subjects in the treatment condition with $S_i=1$ are a mixture of the always survive and $\sti=1;\;S_{Ci}=0$ stratum.
This scenario is broadly similar to the strong monotonicity, one-way noncompliance situation discussed in this paper.
On the other hand, when no monotonicity assumption holds, both $\st$ and $S_C$ will have to be imputed for every subject; further research is necessary to determine the appropriate way to do so.

Another direction of extension involved the principal score model~\eqref{eq:pscore}.
The performance of \textsc{geepers} in the simulation study of Section~\ref{sec:simulation} depended heavily on the factor $\alpha$, which controlled the extent to which covariates could predict $S$.
That suggests that when covariates are high-dimensional, \textsc{geepers}' performance in applications could be optimized with a high-dimensional semi- or non-parametric model.
If so, several further questions emerge.
First, can the parameter $\alpha$ be extended to a more general parameter measuring the prediction accuracy of a non-parametric model?
Second, if the principal score model cannot be formulated as the solution to a set of estimating equations, how should the standard error be computed?
Lastly, can over-fit principal score models cause bias or other estimation problems, and if so, are there ways to protect against overfitting?

\textsc{geepers} is a flexible, easily-implementable method for principal effect estimation for one-way noncompliance, with predictive covariates; extensions to a broader set of circumstances could be a boon to causal modeling.

\section*{Replication Materials}
\sloppy
Replication materials can be found at
\url{https://osf.io/fu2sv/?view_only=b2baebc2f3054ef796717863e4a9662f}.

\section*{AI Disclosure}
We used Grammarly Pro (Version 14.1243.0) and ChatGPT (GPT-4-turbo, current as of July 2025) for editing and language improvement.

\bibliographystyle{apalike}
\bibliography{MOM}

@Manual{PStrata,
    title = {PStrata: Principal Stratification Analysis in R},
    author = {Bo Liu},
    year = {2023},
    note = {R package version 0.0.5},
    url = {https://CRAN.R-project.org/package=PStrata},
    doi = {10.32614/CRAN.package.PStrata},
  }

@manual{rcite,
	title        = {R: A Language and Environment for Statistical Computing},
	author       = {{R Core Team}},
	year         = 2020,
	address      = {Vienna, Austria},
	url          = {https://www.R-project.org/},
	organization = {R Foundation for Statistical Computing}
}

@misc{rstan,
	title        = {{RStan}: the {R} interface to {Stan}},
	author       = {{Stan Development Team}},
	year         = 2020,
	url          = {http://mc-stan.org/},
	note         = {R package version 2.21.2}
}

@article{sandwich,
	title        = {Object-Oriented Computation of Sandwich Estimators},
	author       = {Achim Zeileis},
	year         = 2006,
	journal      = {Journal of Statistical Software},
	volume       = 16,
	number       = 9,
	pages        = {1--16},
	doi          = {10.18637/jss.v016.i09}
}

@inproceedings{aic,
	title        = {Information theory and an extension of the maximum likelihood principle},
	author       = {Akaike, H},
	year         = 1973,
	booktitle    = {2nd International Symposium on Information Theory},
	pages        = {267--281},
	organization = {Akad{\'e}miai Kiad{\'o} Location Budapest, Hungary}
}

@manual{arm,
	title        = {arm: Data Analysis Using Regression and Multilevel/Hierarchical Models},
	author       = {Andrew Gelman and Yu-Sung Su},
	year         = 2022,
	url          = {https://CRAN.R-project.org/package=arm},
	note         = {R package version 1.13-1}
}

@article{air,
	title        = {Identification of causal effects using instrumental variables},
	author       = {Angrist, Joshua D and Imbens, Guido W and Rubin, Donald B},
	year         = 1996,
	journal      = {Journal of the American statistical Association},
	publisher    = {Taylor \& Francis},
	volume       = 91,
	number       = 434,
	pages        = {444--455}
}

@article{fellerEtAl2016,
	title        = {{Compared to what? Variation in the impacts of early childhood education by alternative care type}},
	author       = {Avi Feller and Todd Grindal and Luke Miratrix and Lindsay C. Page},
	year         = 2016,
	journal      = {The Annals of Applied Statistics},
	publisher    = {Institute of Mathematical Statistics},
	volume       = 10,
	number       = 3,
	pages        = {1245 -- 1285},
	doi          = {10.1214/16-AOAS910},
	url          = {https://doi.org/10.1214/16-AOAS910}
}

@book{boosStefanskiBook,
	title={Essential statistical inference: theory and methods},
  author={Boos, Dennis D and Stefanski, Leonard A},
  volume={591},
  year={2013},
  publisher={Springer}
}

@article{bradley1997use,
	title        = {The use of the area under the ROC curve in the evaluation of machine learning algorithms},
	author       = {Bradley, Andrew P},
	year         = 1997,
	journal      = {Pattern recognition},
	publisher    = {Elsevier},
	volume       = 30,
	number       = 7,
	pages        = {1145--1159}
}

@article{growthMindsetRuralBurnette,
	title        = {An online growth mindset intervention in a sample of rural adolescent girls},
	author       = {Burnette, Jeni L and Russell, Michelle V and Hoyt, Crystal L and Orvidas, Kasey and Widman, Laura},
	year         = 2018,
	journal      = {British Journal of Educational Psychology},
	publisher    = {Wiley Online Library},
	volume       = 88,
	number       = 3,
	pages        = {428--445}
}

@book{carroll2006measurement,
	title        = {Measurement error in nonlinear models: a modern perspective},
	author       = {Carroll, Raymond J and Ruppert, David and Stefanski, Leonard A and Crainiceanu, Ciprian M},
	year         = 2006,
	publisher    = {Chapman and Hall/CRC}
}

@article{impactPaper,
	author = {Lauren E. Decker-Woodrow and Craig A. Mason and Ji-Eun Lee and Jenny Yun-Chen Chan and Adam Sales and Allison Liu and Shihfen Tu},
title ={The Impacts of Three Educational Technologies on Algebraic Understanding in the Context of COVID-19},

journal = {AERA Open},
volume = {9},
number = {},
 pages = {23328584231165919},
year = {2023},
doi = {10.1177/23328584231165919},

URL = { 
    
        https://doi.org/10.1177/23328584231165919
    
    

},
eprint = { 
    
        https://doi.org/10.1177/23328584231165919
    
    

}
}

@Book{AER,
    title = {Applied Econometrics with {R}},
    author = {Christian Kleiber and Achim Zeileis},
    year = {2008},
    publisher = {Springer-Verlag},
    address = {New York},
    doi = {10.1007/978-0-387-77318-6},
    url = {https://CRAN.R-project.org/package=AER},
}

@article{dingLu,
	title        = {Principal stratification analysis using principal scores},
	author       = {Ding, P. and Lu, J.},
	year         = 2017,
	journal      = {Journal of the Royal Statistical Society: Series B (Statistical Methodology)},
	volume       = 79,
	pages        = {757--777}
}

@article{ding2011,
	title        = {Identifiability and estimation of causal effects by principal stratification with outcomes truncated by death},
	author       = {Ding, Peng and Geng, Zhi and Yan, Wei and Zhou, Xiao-Hua},
	year         = 2011,
	journal      = {Journal of the American Statistical Association},
	publisher    = {Taylor \& Francis},
	volume       = 106,
	number       = 496,
	pages        = {1578--1591}
}

@article{feller2017principal,
	title        = {Principal score methods: Assumptions, extensions, and practical considerations},
	author       = {Feller, Avi and Mealli, Fabrizia and Miratrix, Luke},
	year         = 2017,
	journal      = {Journal of Educational and Behavioral Statistics},
	publisher    = {Sage Publications Sage CA: Los Angeles, CA},
	volume       = 42,
	number       = 6,
	pages        = {726--758}
}

@article{frangakis,
	title        = {Principal stratification in causal inference},
	author       = {Frangakis, Constantine E and Rubin, Donald B},
	year         = 2002,
	journal      = {Biometrics},
	publisher    = {Wiley Online Library},
	volume       = 58,
	number       = 1,
	pages        = {21--29}
}

@article{griffin2008application,
	title        = {An application of principal stratification to control for institutionalization at follow-up in studies of substance abuse treatment programs},
	author       = {Griffin, Beth Ann and McCaffery, Daniel F and Morral, Andrew R},
	year         = 2008,
	journal      = {The annals of applied statistics},
	publisher    = {NIH Public Access},
	volume       = 2,
	number       = 3,
	pages        = 1034
}

@article{hedgesHedberg,
	title        = {Intraclass correlations and covariate outcome correlations for planning two-and three-level cluster-randomized experiments in education},
	author       = {Hedges, Larry V and Hedberg, Eric C},
	year         = 2013,
	journal      = {Evaluation review},
	publisher    = {Sage Publications Sage CA: Los Angeles, CA},
	volume       = 37,
	number       = 6,
	pages        = {445--489}
}

@inproceedings{commonTrends,
  title={Exploring common trends in online educational experiments},
  author={Prihar, Ethan and Syed, Manaal and Ostrow, Korinn and Shaw, Stacy and Sales, Adam and Heffernan, Neil},
  booktitle={Proceedings of the 15th International Conference on Educational Data Mining},
  year={2022}
}

@inproceedings{feller2016principal,
	title        = {Weak Separation in Mixture Models and Implications for Principal Stratification},
	author       = {Ho, Nhat and Feller, Avi and Greif, Evan and Miratrix, Luke and Pillai, Natesh},
	year         = 2022,
	month        = {28--30 Mar},
	booktitle    = {Proceedings of The 25th International Conference on Artificial Intelligence and Statistics},
	publisher    = {PMLR},
	series       = {Proceedings of Machine Learning Research},
	volume       = 151,
	pages        = {5416--5458},
	url          = {https://proceedings.mlr.press/v151/ho22b.html},
	editor       = {Camps-Valls, Gustau and Ruiz, Francisco J. R. and Valera, Isabel}
}

@article{imbens1997bayesian,
	title        = {Bayesian inference for causal effects in randomized experiments with noncompliance},
	author       = {Imbens, Guido W and Rubin, Donald B},
	year         = 1997,
	journal      = {The annals of statistics},
	publisher    = {JSTOR},
	pages        = {305--327}
}

@article{jiangDing2021,
	title        = {Identification of causal effects within principal strata using auxiliary variables},
	author       = {Jiang, Zhichao and Ding, Peng},
	year         = 2021,
	journal      = {Statistical Science},
	publisher    = {Institute of Mathematical Statistics},
	volume       = 36,
	number       = 4,
	pages        = {493--508}
}

@article{jo,
	title        = {On the use of propensity scores in principal causal effect estimation},
	author       = {Jo, Booil and Stuart, Elizabeth A},
	year         = 2009,
	journal      = {Statistics in medicine},
	publisher    = {Wiley Online Library},
	volume       = 28,
	number       = 23,
	pages        = {2857--2875}
}

@article{jo2002,
	title        = {Estimation of intervention effects with noncompliance: Alternative model specifications},
	author       = {Jo, Booil},
	year         = 2002,
	journal      = {Journal of Educational and Behavioral Statistics},
	publisher    = {Sage Publications Sage CA: Los Angeles, CA},
	volume       = 27,
	number       = 4,
	pages        = {385--409}
}

@article{Barnard01062003,
	title        = {Principal Stratification Approach to Broken Randomized Experiments},
	author       = {John Barnard, Constantine E Frangakis, Jennifer L Hill and Donald B Rubin},
	year         = 2003,
	journal      = {Journal of the American Statistical Association},
	publisher    = {ASA Website},
	volume       = 98,
	number       = 462,
	pages        = {299--323},
	doi          = {10.1198/016214503000071}
}

@article{li2010bayesian,
	title        = {A Bayesian approach to surrogacy assessment using principal stratification in clinical trials},
	author       = {Li, Yun and Taylor, Jeremy MG and Elliott, Michael R},
	year         = 2010,
	journal      = {Biometrics},
	publisher    = {Wiley Online Library},
	volume       = 66,
	number       = 2,
	pages        = {523--531}
}

@article{bounding,
	title        = {Bounding, An Accessible Method for Estimating Principal Causal Effects, Examined and Explained},
	author       = {Luke Miratrix and Jane Furey and Avi Feller and Todd Grindal and Lindsay C. Page},
	year         = 2018,
	journal      = {Journal of Research on Educational Effectiveness},
	publisher    = {Routledge},
	volume       = 11,
	number       = 1,
	pages        = {133--162},
	doi          = {10.1080/19345747.2017.1379576},
	url          = {https://doi.org/10.1080/19345747.2017.1379576},
	eprint       = {https://doi.org/10.1080/19345747.2017.1379576}
}

@article{mealli2004analyzing,
	title        = {Analyzing a randomized trial on breast self-examination with noncompliance and missing outcomes},
	author       = {Mealli, Fabrizia and Imbens, Guido W and Ferro, Salvatore and Biggeri, Annibale},
	year         = 2004,
	journal      = {Biostatistics},
	publisher    = {Oxford University Press},
	volume       = 5,
	number       = 2,
	pages        = {207--222}
}

@article{nolen2011randomization,
	title        = {Randomization-based inference within principal strata},
	author       = {Nolen, Tracy L and Hudgens, Michael G},
	year         = 2011,
	journal      = {Journal of the American Statistical Association},
	publisher    = {Taylor \& Francis},
	volume       = 106,
	number       = 494,
	pages        = {581--593}
}

@article{lidsayPage,
	title        = {Principal stratification as a framework for investigating mediational processes in experimental settings},
	author       = {Page, Lindsay C},
	year         = 2012,
	journal      = {Journal of Research on Educational Effectiveness},
	publisher    = {Taylor \& Francis},
	volume       = 5,
	number       = 3,
	pages        = {215--244}
}

@article{richardson2023estimating,
	title        = {Estimating the effect of a treatment when there is nonadherence in a trial},
	author       = {Richardson, David B and Dukes, Oliver and Tchetgen Tchetgen, Eric J},
	year         = 2023,
	journal      = {American Journal of Epidemiology},
	publisher    = {Oxford University Press},
	volume       = 192,
	number       = 10,
	pages        = {1772--1780}
}

@article{rosenbaum1983central,
	title        = {The central role of the propensity score in observational studies for causal effects},
	author       = {Rosenbaum, Paul R and Rubin, Donald B},
	year         = 1983,
	journal      = {Biometrika},
	publisher    = {Oxford University Press},
	volume       = 70,
	number       = 1,
	pages        = {41--55}
}

@article{mcguire,
  title={Counterintuitive effects of online feedback in middle school math: results from a randomized controlled trial in ASSISTments},
  author={McGuire, Patrick and Tu, Shihfen and Logue, Mary Ellin and Mason, Craig A and Ostrow, Korinn},
  journal={Educational Media International},
  volume={54},
  number={3},
  pages={231--244},
  year={2017},
  publisher={Taylor \& Francis}
}

@article{splawa1990application,
	title        = {On the application of probability theory to agricultural experiments. Essay on principles. Section 9.},
	author       = {Splawa-Neyman, Jerzy},
	year         = 1990,
	journal      = {Statistical Science},
	publisher    = {JSTOR},
	note = {Dabrowska, Dorota M and Speed, TP, Trans. Original work published 1923},
	pages        = {465--472}
}

@article{stefanskiBoos,
	title        = {The calculus of M-estimation},
	author       = {Stefanski, Leonard A and Boos, Dennis D},
	year         = 2002,
	journal      = {The American Statistician},
	publisher    = {Taylor \& Francis},
	volume       = 56,
	number       = 1,
	pages        = {29--38}
}

@article{zhangRubin,
	title        = {Estimation of causal effects via principal stratification when some outcomes are truncated by ``death''},
	author       = {Zhang, Junni L and Rubin, Donald B},
	year         = 2003,
	journal      = {Journal of Educational and Behavioral Statistics},
	publisher    = {Sage Publications Sage CA: Los Angeles, CA},
	volume       = 28,
	number       = 4,
	pages        = {353--368}
}

\appendix

\section{Appendix: Proofs and Calculations}
\subsection{Proof for Lemma \ref{lemma:expectation}}

As a preliminary, note that

\begin{align*}
  \EE[Y_C|\pp]&=\EE\left\{\EE[Y_C|\pp,\st]|\pp\right\}\\
             &=\EE\left\{\EE[Y_C|\st]|\pp\right\}\tag*{by \eqref{eq:assumption}}\\
             &=\EE[\muc1\st+\muc0(1-\st)|\pp]\\
             &=\muc1\pp+\muc0(1-\pp)
\end{align*}

Then we have
\begin{equation*}
  \begin{split}
    \EE[Y_C]=&\EE\EE[Y_C|\pp]=\muc1\EE\pp+\muc0(1-\EE\pp)\\
    =&\muc0+\EE\pp(\muc1-\muc0)
    \end{split}
\end{equation*}

Next we have

\begin{align*}
  \EE[Y_C\pp]&=\EE\left\{\EE[Y_C\pp|\pp]\right\}\\
            &=\EE\left\{\pp\EE[Y_C|\pp]\right\}\\
            &=\EE\left\{\pp\left[\muc1\pp+\muc0(1-\pp)\right]\right\}\\
            &=\EE[\pp]\muc0+\EE[\pp^2](\muc1-\muc0)
\end{align*}

In the treatment group, $\st$ is observed, so
\begin{align*}
    \EE[Y_T]=&\mut0+\EE[\st](\mut1-\mut0)\tag*{and}\\
    \EE[\st Y_T]=&\EE[\st]\mut0+\EE[\st^2](\mut1-\mut0)
\end{align*}

Due to Assumption \ref{ass:rand} (randomization), $\EE[Y|Z=0]=\EE[Y_C]$, $\EE[Y|Z=1]=\EE[Y_T]$, $\EE[Y\pp|Z=0]=\EE[Y_C\pp]$ and $\EE[YS|Z=1]=\EE[Y_T\st]$, completing the proof.

\subsection{Proof for Proposition \ref{prop:reg1}}

Replacing $\sti$ and $\pp$ in \eqref{eq:estEq0} with $\ri$, as in \eqref{eq:ri}, and replacing $\tilde{\Psi}_i$ with $\Psi_i=\begin{psmallmatrix} 1 & 0&1&0\\ 0&1&0&1\\0&0&1&0\\0&0&0&1\end{psmallmatrix}\tilde{\Psi}_i$ gives an equivalent set of estimating equations $\sum_{i=1}^{n_C}\Psi_i=\bm{0}$ with $\Psi_i=$
\begin{equation}\label{eq:estEq1}
\begin{pmatrix}
    Y_i-\muc0-\ri(\muc1-\muc0)-Z_i(\mut0-\muc0)-Z_i\ri(\mut1-\mut0-\muc1+\muc0)\\
    \ri Y_i-\ri\muc0-\ri^2(\muc1-\muc0)-Z_i\ri(\mut0-\muc0)-Z_i\ri^2(\mut0-\mut1-\muc0+\muc1)\\
    Z_iY_i-Z_i\mut0-Z_i\ri (\mut1-\mut0)\\
    Z_i\ri Y_i -Z_i\ri\mut0-Z_i\ri^2(\mut1-\mut0)

\end{pmatrix}
\end{equation}
These are equivalent to the estimating equations for OLS model \eqref{eq:regression0} with $\beta_0=\muc0$, $\beta_1=\muc1-\muc0$, $\beta_2=\mut0-\muc0$, and $\beta_3=\mut1-\mut0-\muc1-\muc0$.
Therefore, under standard OLS regularity conditions the estimated parameter vector $\bm{\hat{\beta}}$ is consistent, completing the proof.

\subsection{A Stronger Version of Proposition \ref{prop:reg2} and a Proof}

\begin{prop}\label{prop:interactions}
  Say, for $i=1,\dots,n$, principal scores $\ppi$ are generated as \eqref{eq:pscore}, with parameters $\bm{\alpha}$ identified and consistently estimable with M-estimation, and there exist $\beta_0$, $\beta_1$, $\beta_2$, $\beta_3$, $\bm{\gamma_1}$, $\bm{\gamma_2}$, $\bm{\gamma_3}$ and $\bm{\gamma_4}$ such that $\{Y_i,Z_i,\sti,\bxy_i\}_{i=1}^n$ are independent and identically distributed with
  \begin{equation}\label{eq:interaction}
    \begin{split}
    \EE[Y_i|\st,Z,\bx]=&\beta_0+\beta_1\sti+\beta_2 Z_i+\beta_3Z_i\sti\\
    &+\bm{\gamma_1}'\bxy_i+\bm{\gamma_2}'\bxy_i\sti+
    \bm{\gamma_3}'\bxy Z_i+\bm{\gamma_4}'\bxy_i Z_i\sti
    \end{split}
  \end{equation}

  Then, under Assumptions \ref{ass:sm}, \ref{ass:rand}, and \ref{ass:vps}, if $\ppi$ is linearly independent of $\bxy$, a researcher may follow the following procedure to estimate principal effects:
  \begin{enumerate}
  \item Estimate principal scores by fitting model \eqref{eq:pscore} to data from the treatment group
  \item Replace $\sti$ with $\ri$ (as defined in \ref{eq:ri}) in model \eqref{eq:interaction} and fit with OLS
  \item Estimate principal effects as:
   \begin{equation}\label{eq:prinEffEstApp}
  \begin{split}
    \heff0_{int}&\equiv \hat{\beta}_2+\bm{\hat{\gamma}_3}'\overline{\bxy}_{Z=1,S=0}\\
    \heff1_{int}&\equiv \hat{\beta}_2+\hat{\beta}_3+(\bm{\hat{\gamma}_3}+\bm{\hat{\gamma}_4})'\overline{\bxy}_{Z=1,S=1}
  \end{split}
   \end{equation}
   where $\overline{\bxy}_{Z=1,S=0}$ and $\overline{\bxy}_{Z=1,S=0}$ are the vector of covariate sample means for the subsets of subjects with $Z=1$ and $S=0$ or $S=1$, respectively.
  \end{enumerate}
  Then $\heff0_{G}$ and $\heff1_{int}$ are M-estimators. If the estimating equations for \eqref{eq:prinEffEstApp} are each bounded by an integrable function of $\{\bm{Y},\bxy, \bm{S},\pp,\bm{Z}\}$ that does not depend on $\{\bm{\beta},\bm{\gamma}\}$, then $\heff0_{int}\rightarrow_p\eff0$ and $\heff1_{int}\rightarrow_p\eff1$ as $n\rightarrow\infty$.

  If the parameter estimates of the principal score model are asymptotically normal, second partial derivatives of the estimating equations for \eqref{eq:prinEffEstApp} are bounded by an integrable function of the data for values of $\{\bm{\beta},\bm{\gamma}\}$ in a neighborhood of their probability limits, and the sandwich components of \eqref{eq:sandwich}, $A$ and $B$, exist and are finite, and if $B$ is non-singular, then $\heff0_{int}$ and $\heff1_{int}$ are jointly asymptotically normal, with a variance of the form \eqref{eq:sandwich}.
\end{prop}

Equation \eqref{eq:interaction} implies Assumption \ref{ass:rci} with $\bm{\gamma_2}=\bm{\gamma_3}=\bm{\gamma_4}=0$.

\begin{proof}
First of all, by \eqref{eq:interaction},
\begin{equation*}
\begin{split}
  \EE[Y_T-Y_C|\st=0]&\\
  =&\EE[Y|Z=1,\st=0]-\EE[Y|Z=1,\st=0]\\
  =&\beta_2+\bm{\gamma_3}'\EE[\bxy|\st=0]
\end{split}
\end{equation*}
and
\begin{equation*}
\begin{split}
  \EE[Y_T-Y_C|\st=1]&\\
  =&\EE[Y|Z=1,\st=1]-\EE[Y|Z=1,\st=1]\\
  =&\beta_2+\beta_3+(\bm{\gamma_3}'+\bm{\gamma_4}')\EE[\bxy|\st=1]
\end{split}
\end{equation*}
Furthermore, $\overline{\bxy}_{Z=1,S=0}\rightarrow \EE[\bxy|\st=0]$ and $\overline{\bxy}_{Z=1,S=1}\rightarrow \EE[\bxy|\st=1]$ as $n\rightarrow \infty$.

We will show that the estimated coefficients from model \eqref{eq:interaction}, but with $R$ replacing $\st$, fit with OLS, are consistent for $\bm{\beta}$ and $\bm{\gamma}$ from \eqref{eq:interaction}.

First, note that $\EE[S]=\EE\EE[S|\bx]=\EE[\pp]$ and
\begin{align*}
  \EE[\bx S]&=\EE[\bx S|Z=1]=\EE[\bx S|Z=0] \mbox{ (due to randomization)}\\
  &=\EE[\bx\EE[S|\bx]|Z=0]=\EE[\bx\pp|Z=0]=\EE[\bx\pp]
\end{align*}
implying that, according to \eqref{eq:interaction},
\begin{equation*}
  \begin{split}
    \EE[Y|\bx,Z=0]&=\beta_0+\beta_1\pp+\bm{\gamma_1}'\bxy+\bm{\gamma_2}'\bxy\pp\\
                  &=\beta_0+\beta_1R+\bm{\gamma_1}'\bxy+\bm{\gamma_2}'\bxy R\\
    \EE[Y|\bx,S,Z=1]&=\beta_0+\beta_2+(\beta_1+\beta_3)S+(\bm{\gamma_1}+\bm{\gamma_3})\bxy+(\bm{\gamma_2}+\bm{\gamma_4})\bxy S\\
    &=\beta_0+\beta_2+(\beta_1+\beta_3)R+(\bm{\gamma_1}+\bm{\gamma_3})\bxy+(\bm{\gamma_2}+\bm{\gamma_4})\bxy R
  \end{split}
\end{equation*}
Therefore,
\begin{align*}
  \EE[Y]=&\EE[Y|Z=0]+\EE[Z]\left\{\EE[Y|Z=1]-\EE[Y|Z=0]\right\}\\
  =&\beta_0+\beta_1\EE[R]+\beta_2\EE[Z]+\beta_3\EE[ZR]+\bm{\gamma_1}'\EE[\bxy]+\bm{\gamma_2}'\EE[\bxy R]\\
  &+\bm{\gamma_3}'\EE[\bxy Z]+\bm{\gamma_4}\EE[\bxy RZ]
\end{align*}

Analogous reasoning leads to expressions for $\EE[RY]$, $\EE[\bxy Y]$, $\EE[ZY]$,  $\EE[ZRY]$, $\EE[\bxy YZ]$, and $\EE[\bxy RZY]$.
These, in turn, give rise to estimating equations
\begin{equation*}
\begin{split}
  &\psi_i=\\ &\begin{pmatrix}
  Y_i\\
  \phantom{}\\
  Y_i\ri\\
    \phantom{}\\
  Y_iZ_i\\
  \phantom{}\\
  Y_iZ_i\ri\\
  \phantom{}\\
  Y_i\bxy_i\\
  \phantom{}\\
  Y_i\bxy_i\ri\\
  \phantom{}\\
  Y_i\bxy_iZ_i\\
  \phantom{}\\
  Y_i\bxy_iZ_i\ri\\
  \phantom{}\end{pmatrix} - \begin{pmatrix*}[l]
    \beta_0+\beta_1R_i+\beta_2Z_i+\beta_3Z_iR_i\\
    \quad+\left\{\bm{\gamma_1}'+R_i\bm{\gamma_2}'+Z_i\bm{\gamma_3}'+R_iZ_i\bm{\gamma_4}'\right\}\bxy\\
    \beta_0\ri+\beta_1R_i^2+\beta_2Z_i\ri+\beta_3Z_iR_i^2\\
    \quad+\ri\left\{\bm{\gamma_1}'+R_i\bm{\gamma_2}'+Z_i\bm{\gamma_3}'+R_iZ_i\bm{\gamma_4}'\right\}\bxy\\
    (\beta_0+\beta_2)Z_i+(\beta_1+\beta_3)R_iZ_i\\
    \quad+Z_i\left\{\bm{\gamma_1}'+R_i\bm{\gamma_2}'+Z_i\bm{\gamma_3}'+R_iZ_i\bm{\gamma_4}'\right\}\bxy\\
    \beta_0Z_i\ri+(\beta_1+\beta_3)Z_iR_i^2+\beta_2Z_i\ri\\
    \quad+Z_iR_i\left\{\bm{\gamma_1}'+\bm{\gamma_3}'+R_i(\bm{\gamma_2}'+\bm{\gamma_4}')\right\}\bxy \\
    \beta_0\bxyt+\beta_1R_i\bxyt+\beta_2Z_i\bxyt+\beta_3Z_iR_i\bxyt\\
    \quad {}+\left\{\bm{\gamma_1}'+R_i\bm{\gamma_2}'+Z_i\bm{\gamma_3}'+R_iZ_i\bm{\gamma_4}'\right\}\bxy\bxyt\\
    \beta_0\ri\bxyt+\beta_1R_i^2\bxyt+\beta_2Z_i\ri\bxyt+\beta_3Z_iR_i^2\bxyt\\
    \quad {}+\ri\left\{\bm{\gamma_1}'+R_i\bm{\gamma_2}'+Z_i\bm{\gamma_3}'+R_iZ_i\bm{\gamma_4}'\right\}\bxy\bxyt\\
        (\beta_0+\beta_2)Z_i\bxyt+(\beta_1+\beta_3)R_iZ_i\bxyt\\
    \quad {}+Z_i\left\{\bm{\gamma_1}'+R_i\bm{\gamma_2}'+Z_i\bm{\gamma_3}'+R_iZ_i\bm{\gamma_4}'\right\}\bxy\bxyt\\
\beta_0\ri Z_i\bxyt+\beta_1R_i^2Z_i\bxyt+\beta_2\ri Z_i\bxyt+\beta_3Z_iR_i^2\bxyt\\
    \quad {}+\ri Z_i\left\{\bm{\gamma_1}'+R_i\bm{\gamma_2}'+Z_i\bm{\gamma_3}'+R_iZ_i\bm{\gamma_4}'\right\}\bxy\bxyt\\
  \end{pmatrix*}
  \end{split}
\end{equation*}
with $\EE[\psi_i]=0$.
These are the estimating equations for the regression model \eqref{eq:interaction}, with $R$ replacing $\st$, fit by OLS.
Consistency and asymptotic normality follow from theorems 7.8.1 and 7.8.2, respectively, of \citet{boosStefanskiBook}
\end{proof}

\subsection{Sandwich Matrix Calculations}

Here we will derive the sandwich variance-covariance matrix for the \textsc{geepers} estimate without interactions between $\bx$ and either $Z$ or $\st$---i.e., with $\bm{\gamma_2}=\bm{\gamma_3}=\bm{\gamma_4}=0$ in the notation of \eqref{eq:interaction}---and estimating principal scores using a generalized linear model.

We propose estimating principal effects in two stages.
First, fit the model
\begin{equation}\label{eq:psMod}
  \ppi=Pr(\sti=1|\bxsi)=f(\bm{\alpha}'\bxsit)
\end{equation}
for some inverse link function $f(\cdot)$, where $\bxsit=[1,\bxsi]$, using (observed) values from the treatment group, and estimating $\hat{\alpha}$.
Then let
\begin{equation}\label{eq:ps}
  \hat{p}_i=f(\bm{\hat{\alpha}}'\bxsit)
\end{equation}
for all subjects in the experiment.

Finally, fit model
\begin{equation}\label{eq:regression}
  Y_i=\beta_0+\beta_1r_i+\beta_2Z_i+\beta_3Z_i\ri+\bm{\gamma}'\bxy_i+\epsilon_i
\end{equation}
to estimate $\bm{\beta}$ and hence principal effects, where $\bm{x}_i$ is a set of covariates predictive of $Y$ within principal strata.
Let $\bm{\beta}=[\beta_0,\beta_1,\beta_2,\beta_3,\bm{\gamma}']'$.

\sloppy
Following \eqref{eq:stacked}, let $\blam(Z_i,S_i,\bxsi,\bxy_i,Y_i;\bm{\alpha},\bm{\beta})=\begin{pmatrix} Z_i\Omega(\bxsi,S_i;\bm{\alpha})\\ \Psi(\bxy_i,\bxsi,Y_i,Z_i,\ri(\bm{\alpha});\bm{\beta})\end{pmatrix}$, the stacked estimating equations of \eqref{eq:psMod} and \eqref{eq:regression}.
Going forward, for the sake of brevity, we will write $\blam_i(\bm{\alpha},\bm{\beta})=\blam(Z_i,S_i,\bxsi,\bxy_i,Y_i;\bm{\hat{\alpha}},\bm{\hat{\beta}})$, where dependence on the data for $i$ is captured in the subscript $i$, with similar meanings for $\Psi_i(\bm{\alpha},\bm{\beta})$ and $\omega_i(\bm{\alpha})$.


The variance-covariance matrix for $\bm{\hat{\alpha}}$ and $\bm{\hat{\beta}}$ can be estimated as:
\begin{equation*}
  \widehat{var}\left([\bm{\hat{\alpha}}',\bm{\hat{\beta}}']'\right)=A^{-1}BA^{-t}
\end{equation*}
where
\begin{equation*}
  A=\sum_i \frac{\partial}{\partial [\bm{\alpha},\bm{\beta}]'} \blam_i\Bigr|_{\substack{\bm{\alpha}=\bm{\hat{\alpha}}\\\bm{\beta}=\bm{\hat{\beta}}}}
\end{equation*}
and
\begin{equation*}
  B=\sum_i \blam_i(\bm{\hat{\alpha}},\bm{\hat{\beta}})\blam_i(\bm{\hat{\alpha}},\bm{\hat{\beta}})'
\end{equation*}

Following \citet[][p. 373]{carroll2006measurement}, we 
can decompose the matrices into diagonal elements
\begin{equation*}
    \begin{split}
        A_{1,1}&=\sum_i \partial \Omega_i/\partial \bm{\alpha}|_{\bm{\alpha}=\bm{\hat{\alpha}}}\\
        A_{2,2}&=\sum_i\partial\Psi_i/\partial \bm{\beta}|_{\bm{\beta}=\bm{\hat{\beta}}}\\
        B_{1,1}&=\sum_i\Omega_i(\bm{\hat\alpha})\Omega_i(\bm{\hat\alpha})'\\
        B_{2,2}&=\sum_i \Psi_i(\bm{\hat\alpha},\bm{\hat\beta})\Psi_i(\bm{\hat\alpha},\bm{\hat\beta})'
    \end{split}
\end{equation*}
 that pertain to the parameter sets $\bm{\alpha}$ and $\bm{\beta}$ and the estimating equations for models \eqref{eq:psMod} and  \eqref{eq:regression}, respectively, and
 \begin{equation*}
     \begin{split}
         A_{21}&=\sum_i\partial\Psi_i/\partial \bm{\alpha}|_{\bm{\alpha}=\bm{\hat{\alpha}}}\\
         B_{12}=B_{21}'&=\sum_i \Omega_i(\bm{\hat{\alpha}})\Psi_i(\bm{\hat{\alpha}},\bm{\hat{\beta}})'
     \end{split}
 \end{equation*}
 which capture the dependence of model \eqref{eq:regression} on the parameters $\bm{\alpha}$ from \eqref{eq:psMod} and the covariance between the estimating equations of the two models.

 The sub-matrix $A_{12}=\sum_i \partial \Omega_i/\partial \bm{\beta}=0$, since \eqref{eq:psMod} does not depend on $\bm{\beta}$.

The diagonal matrices $A_{1,1}$ and $A_{2,2}$ and $B_{1,1}$ and $B_{2,2}$ are all the typical ``bread'' and ``meat'' matrices from M-estimation of generalized linear models and OLS.
Calculation of the matrices $B_{12}$ and $B_{21}$ is straightforward after vectors $\Omega_i(\bm{\hat{\alpha}})$ and $\Psi_i(\bm{\hat{\alpha}},\bm{\hat{\beta}})$ have been calculated.
Some specialized calculation is necessary for matrix $A_{21}$.

\subsection{$A_{21}$ Matrix}
The estimating equations for the regression \eqref{eq:regression} are
\begin{equation}\label{eq:eeOLS}
  \psi(Y_i,\bxy_i,\bxsi,\bm{\beta},\alpha)=X_iY_i-X_iX_i'\bm{\beta}
\end{equation}
Where $X_i=[1,r_i,Z_i,r_iZ_i,\bxyt_i]'$.
In other words,
\begin{align*}
  \psi(Y_i,&\bxy_i,\bxsi,\bm{\beta},\bm{\alpha})=\\
  &\left\{
  \begin{array}{l}
    Y_i-\left(\beta_0+\beta_1r_i+\beta_2Z_i+\beta_3Z_ir_i+\bm{\gamma}'\bxy_i\right)\\
    r_iY_i-r_i\left(\beta_0+\beta_2Z_i+\bm{\gamma}'\bxy_i\right)-r_i^2\left(\beta_1+\beta_3Z_i\right)\\
    Z_iY_i-Z_i\left(\beta_0+\beta_1r_i+\beta_2Z_i+\beta_3Z_ir_i+\bm{\gamma}'\bxy_i\right)\\
    Z_ir_iY_i-Z_ir_i\left(\beta_0+\beta_2Z_i+\bm{\gamma}'\bxy_i\right)-Z_ir_i^2\left(\beta_1+\beta_3Z_i\right)\\
    \bxy_iY_i-\bxy_i\left(\beta_0+\beta_1r_i+\beta_2Z_i+\beta_3Z_ir_i+\bm{\gamma}'\bxy_i\right)
  \end{array}
  \right\}
\end{align*}
(noting that $Z^2=Z$).
These depend on $\bm{\alpha}$ when $r_i=p_i$, i.e., when $Z_i=0$.
Then note that, following \eqref{eq:ps}, and letting $\eta_i=\bm{\alpha}'\bxsit$
\begin{equation}\label{eq:derivP}
  \frac{\partial p_i}{\partial \bm{\alpha}'}=f'(\eta_i)\frac{\partial \eta_i}{\partial \bm{\alpha}'}=f'(\eta_i)\bxsitp
\end{equation}
and that
\begin{equation}
  \frac{\partial p_i^2}{\partial \bm{\alpha}'}=2p_i\frac{\partial p_i}{\partial \bm{\alpha}'}=2f(\eta)f'(\eta_i)\bxsitp=2p_if'(\eta_i)\bxsitp
\end{equation}

Then if $r_i=p_i$,

\begin{align*}
  \frac{\partial}{\partial \bm{\alpha}'}&\psi(Y_i,\bxy_i,\bm{\beta},\bm{\alpha})=\\
  & \left[\begin{array}{c}
          -(\beta_1)f'(\eta_i)\bxsitp\\
          \left[Y_i-X_i'\bm{\beta}-2p_i(\beta_1)\right]f'(\eta_i)\bxsitp\\
          0\\
          0\\
          -\beta_1f'(\eta_i)\bxy_i\bxsitp
    \end{array}\right]
\end{align*}

If $r_i=S_i$, $\frac{\partial}{\partial \bm{\alpha}'}\psi(Y_i,\bxy_i,\bm{\beta},\bm{\alpha})=0$.

  \FloatBarrier
\section{Additional Simulation Results}
\FloatBarrier
\subsection{Plot of  AUC versus $\alpha$}
\FloatBarrier

\begin{center}

    \includegraphics[width=5.5in]{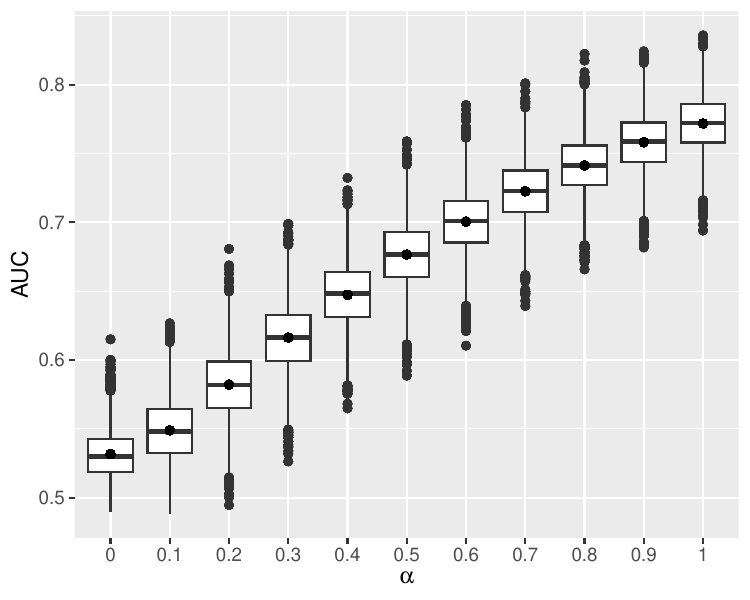}

\end{center}

\FloatBarrier
\subsection{Full Empirical 95\% Interval Coverage Results}
\FloatBarrier

The following tables give the empirical coverage of nominal 95\% intervals for \textsc{geepers} and mixture model principal effect estimates under varying data generating models. 

\begin{table}
  \caption{Empirical coverage of nominal 95\% Confidence intervals for \geepers and \pmm when $n=500$ per condition.}
  
\begin{tabular}[t]{lllrlllllll}
\toprule
\multicolumn{5}{c}{ } & \multicolumn{6}{c}{$n=500$} \\
\cmidrule(l{3pt}r{3pt}){6-11}
\multicolumn{5}{c}{ } & \multicolumn{2}{c}{$\alpha=0$} & \multicolumn{2}{c}{$\alpha=0.3$} & \multicolumn{2}{c}{$\alpha=0.5$} \\
\cmidrule(l{3pt}r{3pt}){6-7} \cmidrule(l{3pt}r{3pt}){8-9} \cmidrule(l{3pt}r{3pt}){10-11}
\makecell[l]{Residual\\Dist.} & \makecell[l]{$\bm{x}:Z$\\Int.?} & \makecell[l]{$\bm{x}:S_T$\\Int.?} & $\beta_1$ & \makecell[l]{Prin.\\Eff} & \textsc{geepers} & \textsc{pmm} & \textsc{geepers} & \textsc{pmm} & \textsc{geepers} & \textsc{pmm}\\
\midrule
 & No & No & 0 & $\tau^0$ & 1.00 & 0.99 & 0.96 & 0.99 & 0.95 & 0.98\\

 & No & No & 0 & $\tau^1$ & 1.00 & 0.99 & 0.97 & 0.99 & 0.95 & 0.97\\

 & Yes & No & 0 & $\tau^0$ & \rd{0.92} & \rd{0.98} & \rd{0.41} & \rd{0.72} & \rd{0.40} & \rd{0.55}\\

 & Yes & No & 0 & $\tau^1$ & \rd{0.92} & \rd{0.98} & \rd{0.41} & \rd{0.69} & \rd{0.40} & \rd{0.53}\\

 & No & Yes & 0 & $\tau^0$ & \rd{1.00} & \rd{0.99} & \rd{0.96} & \rd{0.99} & \rd{0.96} & \rd{0.98}\\

 & No & Yes & 0 & $\tau^1$ & \rd{1.00} & \rd{0.99} & \rd{0.96} & \rd{0.99} & \rd{0.95} & \rd{0.98}\\

 & Yes & Yes & 0 & $\tau^0$ & \rd{0.92} & \rd{0.98} & \rd{0.41} & \rd{0.72} & \rd{0.40} & \rd{0.57}\\

\multirow{-8}{*}{\raggedright\arraybackslash Lognormal} & Yes & Yes & 0 & $\tau^1$ & \rd{0.92} & \rd{0.98} & \rd{0.42} & \rd{0.69} & \rd{0.42} & \rd{0.54}\\
\cmidrule{1-11}
 & No & No & 0 & $\tau^0$ & 1.00 & 1.00 & 0.96 & 0.96 & 0.96 & 0.95\\

 & No & No & 0 & $\tau^1$ & 1.00 & 1.00 & 0.96 & 0.97 & 0.96 & 0.95\\

 & Yes & No & 0 & $\tau^0$ & \rd{0.92} & \rd{0.89} & \rd{0.41} & \rd{0.37} & \rd{0.39} & \rd{0.33}\\

 & Yes & No & 0 & $\tau^1$ & \rd{0.91} & \rd{0.89} & \rd{0.41} & \rd{0.37} & \rd{0.39} & \rd{0.34}\\

 & No & Yes & 0 & $\tau^0$ & \rd{1.00} & \rd{1.00} & \rd{0.97} & \rd{0.96} & \rd{0.96} & \rd{0.95}\\

 & No & Yes & 0 & $\tau^1$ & \rd{1.00} & \rd{1.00} & \rd{0.96} & \rd{0.96} & \rd{0.96} & \rd{0.95}\\

 & Yes & Yes & 0 & $\tau^0$ & \rd{0.92} & \rd{0.91} & \rd{0.42} & \rd{0.38} & \rd{0.41} & \rd{0.36}\\

\multirow{-8}{*}{\raggedright\arraybackslash Normal} & Yes & Yes & 0 & $\tau^1$ & \rd{0.92} & \rd{0.89} & \rd{0.42} & \rd{0.37} & \rd{0.42} & \rd{0.35}\\
\cmidrule{1-11}
 & No & No & 0 & $\tau^0$ & 0.99 & \rd{0.25} & 0.94 & \rd{0.27} & 0.96 & \rd{0.47}\\

 & No & No & 0 & $\tau^1$ & 0.99 & \rd{0.25} & 0.94 & \rd{0.27} & 0.95 & \rd{0.47}\\

 & Yes & No & 0 & $\tau^0$ & \rd{0.85} & \rd{0.12} & \rd{0.34} & \rd{0.05} & \rd{0.36} & \rd{0.11}\\

 & Yes & No & 0 & $\tau^1$ & \rd{0.85} & \rd{0.12} & \rd{0.33} & \rd{0.05} & \rd{0.36} & \rd{0.11}\\

 & No & Yes & 0 & $\tau^0$ & \rd{0.99} & \rd{0.42} & \rd{0.96} & \rd{0.38} & \rd{0.96} & \rd{0.54}\\

 & No & Yes & 0 & $\tau^1$ & \rd{1.00} & \rd{0.42} & \rd{0.96} & \rd{0.38} & \rd{0.95} & \rd{0.54}\\

 & Yes & Yes & 0 & $\tau^0$ & \rd{0.84} & \rd{0.21} & \rd{0.36} & \rd{0.08} & \rd{0.38} & \rd{0.15}\\

\multirow{-8}{*}{\raggedright\arraybackslash Uniform} & Yes & Yes & 0 & $\tau^1$ & \rd{0.84} & \rd{0.21} & \rd{0.37} & \rd{0.08} & \rd{0.39} & \rd{0.14}\\
\bottomrule
\end{tabular}

 \end{table}



The following table gives the root mean squared error (RMSE), $\left\{\sum_b (\hat{\tau}-\tau)^2/500\right\}^{1/2}$, for \textsc{geepers}, mixture model, and principal score weighting principal effect estimates under varying data generating models.\\

\begin{table}
  \caption{Empirical RMSE  for \geepers, \pmm, and \psw when $n=500$ per condition.}
  
\begin{tabular}[t]{llllrrrrrr}
\toprule
\multicolumn{4}{c}{ } & \multicolumn{3}{c}{$\alpha=0.3$} & \multicolumn{3}{c}{$\alpha=0.5$} \\
\cmidrule(l{3pt}r{3pt}){5-7} \cmidrule(l{3pt}r{3pt}){8-10}
\makecell[l]{Residual\\Dist.} & \makecell[l]{$\bm{x}:Z$\\Int.?} & \makecell[l]{$\bm{x}:S_T$\\Int.?} & \makecell[l]{Prin.\\Eff} & \textsc{geepers} & \textsc{pmm} & \textsc{psw} & \textsc{geepers} & \textsc{pmm} & \textsc{psw}\\
\midrule
 & No & No & $\tau^0$ & 0.28 & 0.12 & 0.09 & 0.18 & 0.11 & 0.11\\

 & No & No & $\tau^1$ & 0.28 & 0.11 & 0.09 & 0.18 & 0.12 & 0.11\\

 & Yes & No & $\tau^0$ & 0.79 & 0.24 & 0.09 & 0.46 & 0.25 & 0.11\\

 & Yes & No & $\tau^1$ & 0.79 & 0.21 & 0.10 & 0.47 & 0.23 & 0.11\\

 & No & Yes & $\tau^0$ & 0.28 & 0.11 & 0.09 & 0.18 & 0.11 & 0.10\\

 & No & Yes & $\tau^1$ & 0.28 & 0.11 & 0.09 & 0.18 & 0.12 & 0.11\\

 & Yes & Yes & $\tau^0$ & 0.79 & 0.24 & 0.09 & 0.46 & 0.25 & 0.11\\

\multirow{-8}{*}{\raggedright\arraybackslash Lognormal} & Yes & Yes & $\tau^1$ & 0.79 & 0.22 & 0.10 & 0.47 & 0.24 & 0.12\\
\cmidrule{1-10}
 & No & No & $\tau^0$ & 0.28 & 0.17 & 0.09 & 0.18 & 0.15 & 0.12\\

 & No & No & $\tau^1$ & 0.28 & 0.17 & 0.09 & 0.18 & 0.15 & 0.11\\

 & Yes & No & $\tau^0$ & 0.78 & 0.37 & 0.10 & 0.47 & 0.36 & 0.12\\

 & Yes & No & $\tau^1$ & 0.78 & 0.37 & 0.09 & 0.47 & 0.36 & 0.12\\

 & No & Yes & $\tau^0$ & 0.28 & 0.17 & 0.09 & 0.18 & 0.15 & 0.11\\

 & No & Yes & $\tau^1$ & 0.28 & 0.17 & 0.10 & 0.18 & 0.15 & 0.12\\

 & Yes & Yes & $\tau^0$ & 0.78 & 0.37 & 0.09 & 0.47 & 0.36 & 0.11\\

\multirow{-8}{*}{\raggedright\arraybackslash Normal} & Yes & Yes & $\tau^1$ & 0.78 & 0.37 & 0.10 & 0.47 & 0.36 & 0.12\\
\cmidrule{1-10}
 & No & No & $\tau^0$ & 0.30 & 0.51 & 0.09 & 0.18 & 0.42 & 0.12\\

 & No & No & $\tau^1$ & 0.30 & 0.51 & 0.09 & 0.18 & 0.42 & 0.11\\

 & Yes & No & $\tau^0$ & 0.80 & 0.60 & 0.10 & 0.48 & 0.57 & 0.12\\

 & Yes & No & $\tau^1$ & 0.79 & 0.59 & 0.10 & 0.48 & 0.57 & 0.12\\

 & No & Yes & $\tau^0$ & 0.29 & 0.49 & 0.09 & 0.18 & 0.40 & 0.11\\

 & No & Yes & $\tau^1$ & 0.29 & 0.48 & 0.10 & 0.19 & 0.40 & 0.12\\

 & Yes & Yes & $\tau^0$ & 0.79 & 0.59 & 0.09 & 0.48 & 0.56 & 0.12\\

\multirow{-8}{*}{\raggedright\arraybackslash Uniform} & Yes & Yes & $\tau^1$ & 0.79 & 0.59 & 0.10 & 0.48 & 0.56 & 0.12\\
\bottomrule
\end{tabular}

 \end{table}



\FloatBarrier
\section{Additional Results from the Video Scaffolding Study}

\subsection{Diagnostic Plots}
\subsubsection{GEEPERS}
\FloatBarrier
\singlespacing
\FloatBarrier

\begin{figure}
\centering
\includegraphics{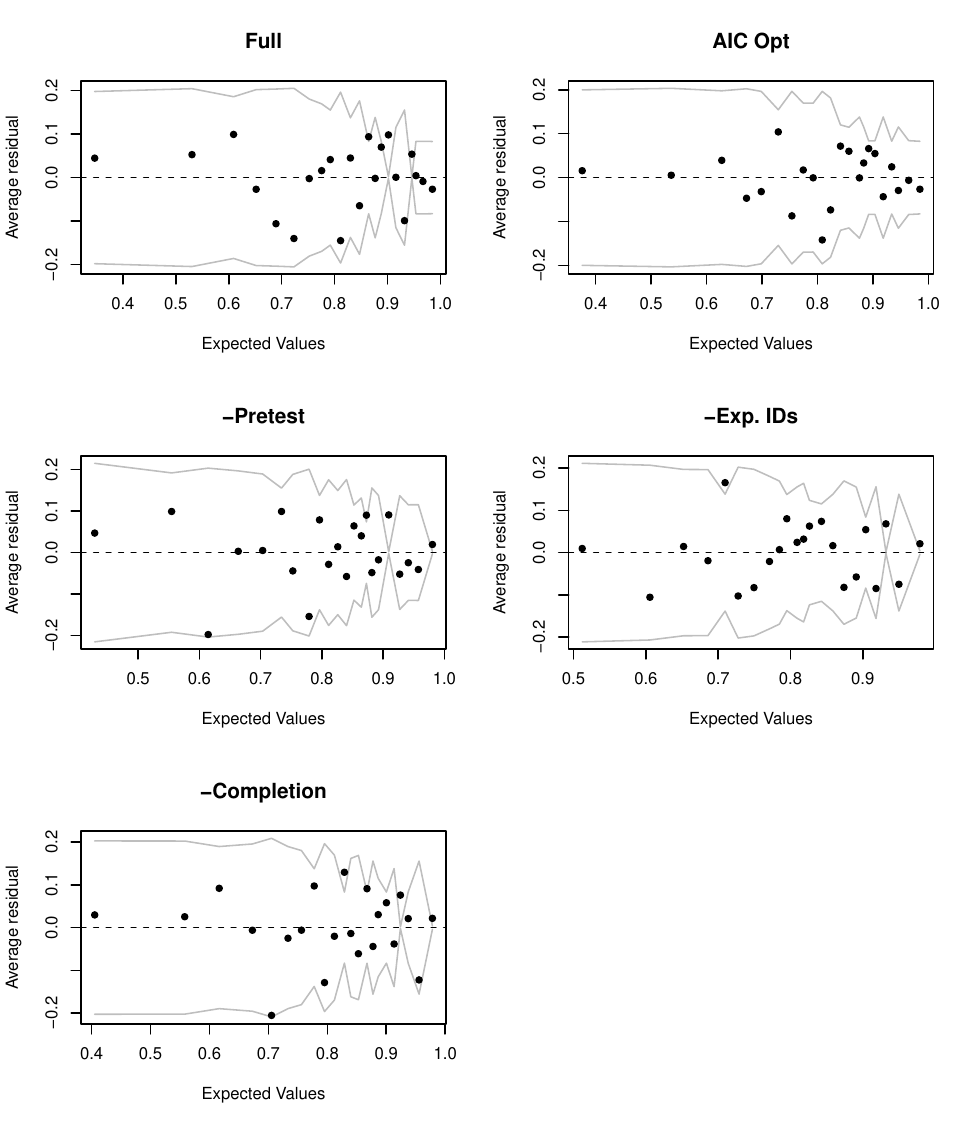}
\caption{Binned residual plots \citep{arm} for the five principal score models in Section \ref{sec:application}.}
\label{fig:binnedResids}
\end{figure}

\begin{figure}
\centering
\includegraphics{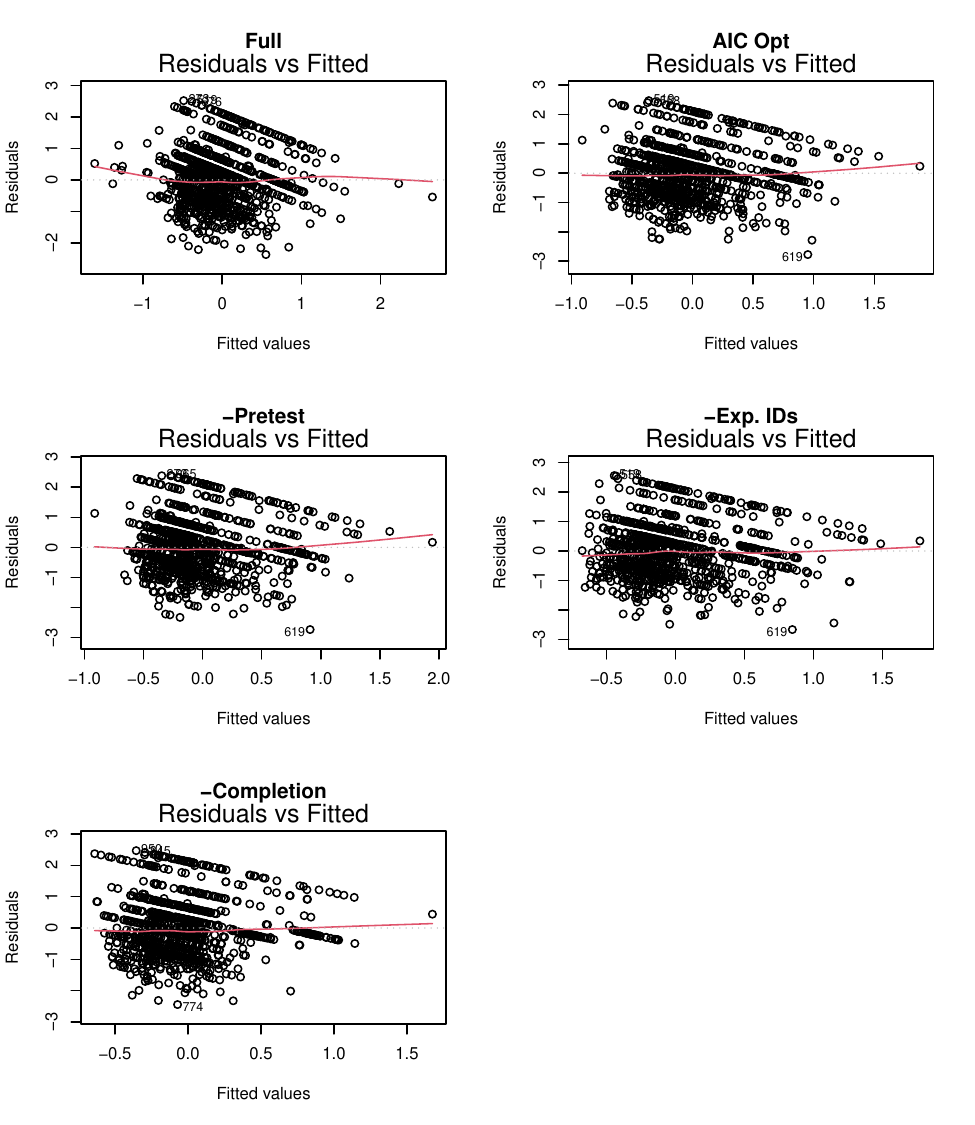}
\caption{Residual versus fitted-value plots for the five \geepers outcome models in Section \ref{sec:application}.}
\label{fig:outResids}
\end{figure}

\FloatBarrier
\subsubsection{Parametric Mixture Model}

\begin{figure}
\centering
\includegraphics{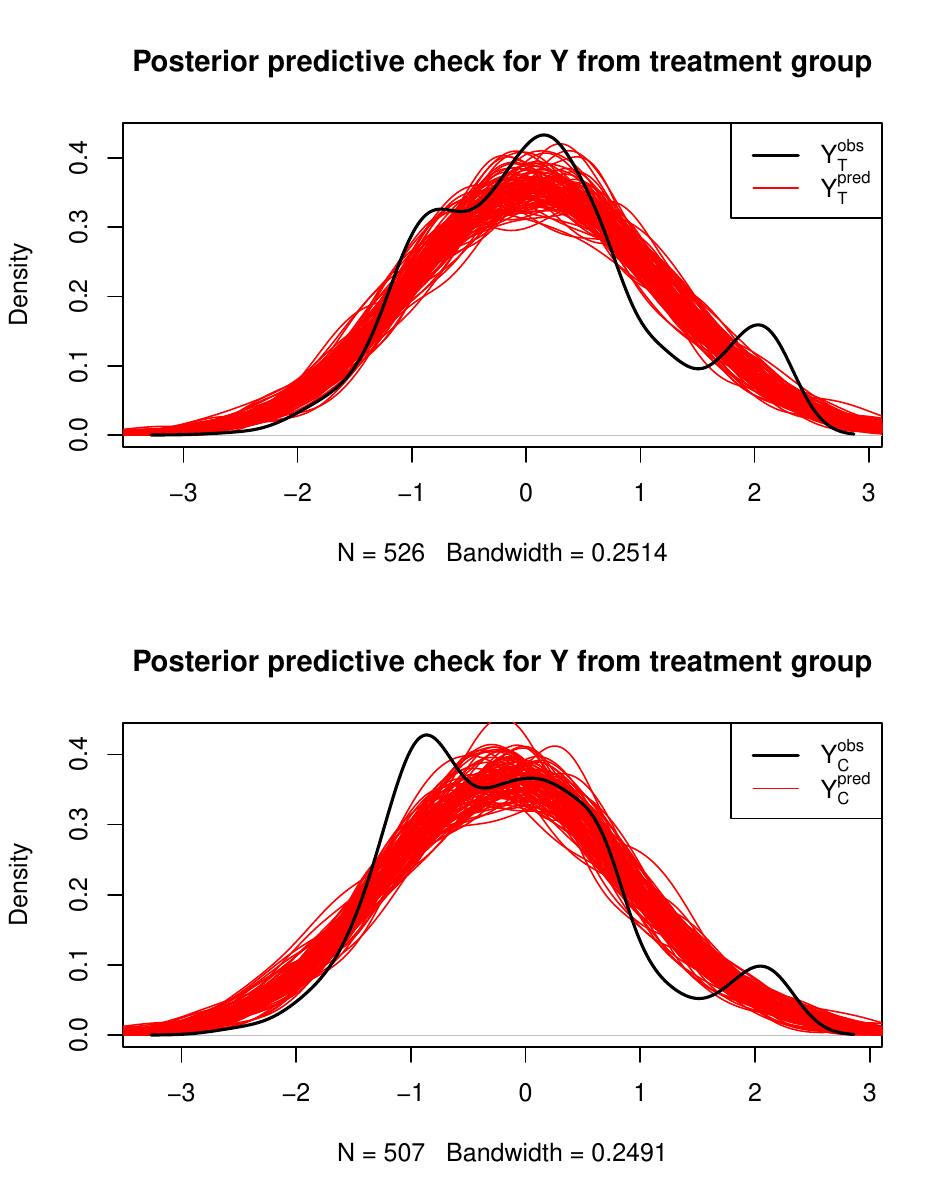}
\caption{Posterior predictive check for the parametric mixture model in Section \ref{sec:application}---density plots for outcomes in the treatment and control groups, along with posterior predictive densities from 100 random posterior draws.}
\label{fig:postPred}
\end{figure}

\FloatBarrier
\subsection{Regression Results}
\FloatBarrier

Regression estimates from five principal score logit models and associated outcome regressions for \textsc{geepers} estimates. Standard errors shown are nominal regression errors, not sandwich corrected. 

\small

\begin{table}
\begin{center}
\begin{tabular}{l c c c c c}
\hline
 & Full & AIC Opt & -Pretest & -Exp. IDs & -Completion \\
\hline
(Intercept)                                        & $2.41 \; (0.29)$  & $2.04 \; (0.18)$  & $1.97 \; (0.17)$  & $1.60 \; (0.13)$  & $1.95 \; (0.17)$  \\
experiment\_idB                              & $-0.14 \; (0.63)$ & $-0.27 \; (0.59)$ & $-0.28 \; (0.59)$ &                   & $-0.23 \; (0.58)$ \\
experiment\_idC                              & $-1.07 \; (0.32)$ & $-1.30 \; (0.27)$ & $-1.14 \; (0.26)$ &                   & $-1.15 \; (0.26)$ \\
stud.\_started\_skill\_builder\_count     & $-0.24 \; (0.30)$ &                   &                   &                   &                   \\
stud.\_skill\_builder\_pct.\_comp. & $0.31 \; (0.16)$  & $0.30 \; (0.15)$  & $0.17 \; (0.14)$  & $0.22 \; (0.14)$  &                   \\
stud.\_started\_problem\_set\_count       & $1.26 \; (0.64)$  & $0.88 \; (0.34)$  & $0.83 \; (0.32)$  & $0.95 \; (0.32)$  & $0.83 \; (0.33)$  \\
stud.\_problem\_set\_pct.\_comp.   & $-0.03 \; (0.14)$ &                   &                   &                   &                   \\
stud.\_comp.\_problem\_count          & $-0.13 \; (0.56)$ &                   &                   &                   &                   \\
stud.\_med.\_first\_resp.\_time      & $0.30 \; (0.34)$  &                   &                   &                   &                   \\
stud.\_med.\_time\_on\_task             & $-0.28 \; (0.35)$ &                   &                   &                   &                   \\
stud.\_avg\_attempt\_count            & $0.38 \; (0.19)$  & $0.32 \; (0.16)$  & $0.53 \; (0.14)$  & $0.49 \; (0.16)$  & $0.29 \; (0.16)$  \\
stud.\_avg\_crct.               & $-0.38 \; (0.19)$ & $-0.46 \; (0.15)$ &                   & $-0.33 \; (0.15)$ & $-0.44 \; (0.14)$ \\
class\_age\_in\_days                               & $0.27 \; (0.20)$  &                   &                   &                   &                   \\
class\_stud.\_count                              & $-0.35 \; (0.16)$ & $-0.20 \; (0.11)$ & $-0.23 \; (0.11)$ & $-0.11 \; (0.11)$ & $-0.20 \; (0.11)$ \\
class\_started\_skill\_builder\_count       & $-0.06 \; (0.25)$ &                   &                   &                   &                   \\
class\_skill\_builder\_pct.\_comp.   & $-0.33 \; (0.20)$ & $-0.40 \; (0.17)$ & $-0.44 \; (0.17)$ & $-0.23 \; (0.15)$ &                   \\
class\_started\_problem\_set\_count         & $-1.41 \; (0.55)$ & $-0.90 \; (0.29)$ & $-0.87 \; (0.28)$ & $-0.88 \; (0.28)$ & $-0.81 \; (0.28)$ \\
class\_problem\_set\_pct.\_comp.     & $-0.22 \; (0.19)$ &                   &                   &                   &                   \\
class\_comp.\_problem\_count            & $0.36 \; (0.39)$  &                   &                   &                   &                   \\
class\_med.\_first\_resp.\_time        & $-1.02 \; (0.90)$ &                   &                   &                   &                   \\
class\_med.\_time\_on\_task               & $0.84 \; (0.88)$  &                   &                   &                   &                   \\
class\_avg\_attempt\_count              & $-0.12 \; (0.20)$ &                   &                   &                   &                   \\
class\_avg\_crct.                 & $-0.08 \; (0.19)$ &                   &                   &                   &                   \\
teacher\_account\_age\_in\_days                    & $-0.00 \; (0.00)$ &                   &                   &                   &                   \\
opportunity\_zone                                  & $0.18 \; (0.15)$  &                   &                   &                   &                   \\
\hline
AIC                                                & $498.84$          & $477.95$          & $485.72$          & $497.07$          & $481.42$          \\
BIC                                                & $605.47$          & $520.60$          & $524.11$          & $531.19$          & $515.54$          \\
Log Likelihood                                     & $-224.42$         & $-228.97$         & $-233.86$         & $-240.53$         & $-232.71$         \\
Deviance                                           & $448.84$          & $457.95$          & $467.72$          & $481.07$          & $465.42$          \\
Num. obs.                                          & $526$             & $526$             & $526$             & $526$             & $526$             \\
\hline
\end{tabular}
\caption{Coefficient estimates for the principal score logistic regressions described in Section \ref{sec:application}}
\label{tab:psMods}
\end{center}
\end{table}

\begin{table}
\begin{center}
\begin{tabular}{l c c c c c}
\hline
 & Full & AIC Opt & -Pretest & -Exp. IDs & -Completion \\
\hline
(Intercept)                                        & $0.96 \; (0.33)$  & $0.91 \; (0.33)$  & $0.65 \; (0.37)$  & $1.34 \; (0.41)$  & $1.47 \; (0.37)$  \\
Z                                                  & $-0.02 \; (0.30)$ & $-0.04 \; (0.32)$ & $0.21 \; (0.35)$  & $-0.62 \; (0.41)$ & $-0.54 \; (0.35)$ \\
R                                                 & $-1.24 \; (0.38)$ & $-1.20 \; (0.40)$ & $-0.88 \; (0.44)$ & $-1.86 \; (0.51)$ & $-1.87 \; (0.44)$ \\
experiment\_idB                              & $0.04 \; (0.12)$  & $-0.06 \; (0.12)$ & $-0.05 \; (0.12)$ &                   & $-0.12 \; (0.12)$ \\
experiment\_idC                              & $-0.41 \; (0.08)$ & $-0.38 \; (0.08)$ & $-0.36 \; (0.08)$ &                   & $-0.43 \; (0.08)$ \\
stud.\_started\_skill\_builder\_count     & $0.06 \; (0.06)$  &                   &                   &                   &                   \\
stud.\_skill\_builder\_pct.\_comp. & $-0.16 \; (0.04)$ & $-0.22 \; (0.04)$ & $-0.22 \; (0.03)$ & $-0.22 \; (0.03)$ &                   \\
stud.\_started\_problem\_set\_count       & $0.59 \; (0.12)$  & $-0.03 \; (0.06)$ & $-0.04 \; (0.06)$ & $0.03 \; (0.06)$  & $-0.02 \; (0.06)$ \\
stud.\_problem\_set\_pct.\_comp.   & $-0.02 \; (0.03)$ &                   &                   &                   &                   \\
stud.\_comp.\_problem\_count          & $-0.58 \; (0.10)$ &                   &                   &                   &                   \\
stud.\_med.\_first\_resp.\_time      & $0.01 \; (0.09)$  &                   &                   &                   &                   \\
stud.\_med.\_time\_on\_task             & $-0.01 \; (0.09)$ &                   &                   &                   &                   \\
stud.\_avg\_attempt\_count            & $-0.03 \; (0.04)$ & $-0.02 \; (0.04)$ & $-0.04 \; (0.03)$ & $0.04 \; (0.04)$  & $0.01 \; (0.04)$  \\
stud.\_avg\_crct.               & $-0.00 \; (0.04)$ & $0.04 \; (0.04)$  &                   & $0.05 \; (0.04)$  & $-0.08 \; (0.04)$ \\
class\_age\_in\_days                               & $0.08 \; (0.04)$  &                   &                   &                   &                   \\
class\_stud.\_count                              & $-0.09 \; (0.04)$ & $-0.00 \; (0.03)$ & $0.00 \; (0.03)$  & $0.02 \; (0.03)$  & $0.01 \; (0.03)$  \\
class\_started\_skill\_builder\_count       & $0.01 \; (0.06)$  &                   &                   &                   &                   \\
class\_skill\_builder\_pct.\_comp.   & $-0.04 \; (0.04)$ & $-0.04 \; (0.03)$ & $-0.03 \; (0.03)$ & $-0.03 \; (0.03)$ &                   \\
class\_started\_problem\_set\_count         & $-0.27 \; (0.12)$ & $0.08 \; (0.06)$  & $0.09 \; (0.06)$  & $0.04 \; (0.06)$  & $0.11 \; (0.06)$  \\
class\_problem\_set\_pct.\_comp.     & $-0.01 \; (0.03)$ &                   &                   &                   &                   \\
class\_comp.\_problem\_count            & $0.20 \; (0.09)$  &                   &                   &                   &                   \\
class\_med.\_first\_resp.\_time        & $-0.19 \; (0.17)$ &                   &                   &                   &                   \\
class\_med.\_time\_on\_task               & $0.15 \; (0.17)$  &                   &                   &                   &                   \\
class\_avg\_attempt\_count              & $0.00 \; (0.05)$  &                   &                   &                   &                   \\
class\_avg\_crct.                 & $0.03 \; (0.04)$  &                   &                   &                   &                   \\
teacher\_account\_age\_in\_days                    & $-0.00 \; (0.00)$ &                   &                   &                   &                   \\
opportunity\_zone                                  & $0.09 \; (0.03)$  &                   &                   &                   &                   \\
Z:R                                               & $0.39 \; (0.37)$  & $0.40 \; (0.39)$  & $0.09 \; (0.43)$  & $1.12 \; (0.50)$  & $1.01 \; (0.43)$  \\
\hline
R$^2$                                              & $0.20$            & $0.15$            & $0.14$            & $0.14$            & $0.11$            \\
Adj. R$^2$                                         & $0.17$            & $0.14$            & $0.13$            & $0.14$            & $0.10$            \\
Num. obs.                                          & $1033$            & $1033$            & $1033$            & $1033$            & $1033$            \\
\hline
\end{tabular}
\caption{Coefficient estimates for the outcome regressions described in Section \ref{sec:application}}
\label{tab:outModAppendix}
\end{center}
\end{table}




\end{document}